\documentclass[seceq,preprint]{jpsj2}
\usepackage{epsfig}
\usepackage{amsmath,amssymb}
\usepackage{latexsym}
\newcommand{\mbold}[1]{\mbox{\boldmath $ #1 $}}

\def\beq{\begin{equation}}
\def\eeq{\end{equation}}
\def\beqa{\begin{eqnarray}}
\def\bmp{\begin{figure}\begin{minipage}{120mm}}
\def\emp{\end{minipage}\end{figure}}

\def\runauthor{Shinichiro  {\sc YOSHIKAWA}, Kouichi  {\sc Okunishi},
Makoto {\sc SENDA}, and Seiji {\sc MIYASHITA}}

\title
{Quantum Fluctuation-Induced Phase Transition in $S=1/2 $XY-like Heisenberg
Antiferromagnets on the Triangular Lattice}

\author {
Shin-ichiro  {\sc Yoshikawa}
\footnote{E-mail address: yoshikawa@spin.t.u-tokyo.ac.jp}
Kouichi  {\sc Okunishi}$^{1)}$
\footnote{E-mail address: okunishi@phys.sc.niigata-u.ac.jp}
 Makoto {\sc Senda}
\footnote{E-mail address: senda@r9.dion.ne.jp}\\
and
Seiji {\sc Miyashita}
\footnote{E-mail address: miya@spin.t.u-tokyo.ac.jp}
}
\inst
{
Department of Applied Physics, The University of Tokyo,
Hongo 7-3-1, Bunkyo-ku, Tokyo 113-8656, Japan
\\
$1)$
Department of Physics, Niigata University,
Niigata 950-2181, Japan
}

\recdate { \today }

\abst {
The selection of the ground state among nearly degenerate states due to
 quantum fluctuations is studied for the $S=1/2$ 
XY-like Heisenberg antiferromagnets
on the triangular lattice in the magnetic field applied along the
hard axis, which was first pointed out by Nikuni and Shiba.
We find that 
the selected ground state sensitively
depends on the degree of the anisotropy and
the magnitude of the magnetic field. This dependence is similar to
that in the corresponding classical model at finite temperatures where
various types of field induced phases appear
due to the entropy effect.
It is also found that the similarity of the selected states 
in the classical and quantum models
are not the case  in a two-leg ladder lattice,
although the lattice consists of triangles locally
and the ground state of this lattice in the classical case
is the same as that of the triangular lattice.
}

\kword
{Frustration, Triangular lattice, Entropy-induced phase,
Quantum fluctuation induced phase, PWFRG, DMRG}

\begin{document}
\sloppy
\maketitle

\section{Introduction}

The ordering process in triangular lattice antiferromagnets has attracted much interest
because the frustration among antiferromagnetic interactions causes
complicated order parameters. In such systems, 
various spin configurations are nearly degenerate. 
Among them one configuration is selected as a result of a very subtle
balance of the energy and the entropy.
At finite temperatures, a structure that is entropically favorable would be chosen even if it is less favorable in energy. In such cases, we
have a phase transition between the entropically favorable ordered phase
and an energetically favorable ordered phase.
Indeed, such fluctuation-induced phase transitions have been studied 
in the magnetization process of antiferromagnets on the triangular lattice.
\cite{K-XY-h,DHLee,KM-H-h,miyashita-IH-h,miyashita-ptp,Henley,
WMS}

A similar selection of the states occurs also for the ground state of a quantum triangular lattice antiferromagnets at zero temperature;
the classical ground state is not necessarily most stable and some other configuration may become the ground state when a contribution of quantum fluctuations to the energy is taken into account. 
Among nearly degenerate classical configurations, a certain state is selected, 
where quantum fluctuations play a role similar to thermal fluctuations in the classical case at finite temperatures.
In particular, such example was pointed out by Nikuni and Shiba\cite{NS1,JNS,NS2,Chubukov} to understand the ground state phase transition of CsCuCl$_3$.
\cite{motokawaetal,WWWLS}
This material has a hexagonal lattice consisting of layers of triangular antiferromagnets with a weak XY anisotropy and ferromagnetic couplings between the layers. 
It has also been pointed out that RbFeCl$_3$ has qualitatively the same structure, and a similar phase diagram has been obtained.
\cite{shiba}

In the classical spin model, the ground state of the
XY-like Heisenberg antiferromagnet on the triangular lattice
consists of the 120$^{\circ}$ structure. This structure of the
ground state causes the so-called chiral phase transition
due to its two-fold discrete degeneracy.\cite{Miyashita-Shiba}
When the field is applied in the $z$-direction, 
the 120$^{\circ}$ structure is folded to be a shape of 
umbrella which is the most energetically favorable in the field.(Fig. 1(a)) 
The magnetization increases proportionally to the field strength up to the saturation. There is no singularity in the magnetization process.
However, a jump of the magnetization was found in the experiment of CsCuCl$_3$. Nikuni and Shiba have pointed out that this jump is due to 
a phase transition in the ground state.
The quantum correction on the ground state energy can cause a ground state phase transition. By comparing energies of the umbrella structure and a coplanar structure depicted in
(Fig. 1(b)) which we shall call `v-shape' structure,\cite{NS1,JNS,NS2}
they found that the umbrella structure is favorable at low field 
while the v-shape structure is favorable at high field. 
That is,  
although the latter has higher energy in the classical system, it has the lowest energy if the quantum correction of the energy is taken into account.

Recently, Watarai et al.\cite{WMS} have pointed out that the role of quantum fluctuations in the above scenario for the quantum case is very similar to that of thermal fluctuations in the corresponding classical spin system at finite temperatures.
Moreover they have shown on the basis of the detailed temperature-field phase diagram of the classical system that the structure of the phase diagrams is sensitive to the degree of the anisotropy.
That is, no phase transition occurs when the XY anisotropy is strong, and the phase transition from the umbrella structure to the v-shape phase occurs when the system has a moderate anisotropy.
In the Heisenberg limit, the system shows a qualitatively different sequence of phases from that of the case of the moderate XY anisotropy.
In the Heisenberg system, the ground state is continuously degenerate\cite{K-XY-h}.
In the low-field region a coplanar structure, which we shall call a `Y-shape' structure (Fig. 1(c)), is selected by the entropy effect.
In the high-field region, the phase of the v-shape structure becomes the most
stable state as in the case of the XY-like Heisenberg model.
Between the low and high field phases, a collinear phase (Fig. 1(d)) is present, which leads to the $1/3$ plateau in the magnetization process.
When the XY anisotropy becomes very weak, a complex sequence of phases appears 
including phases in both of the above two cases.
Then, it is an interesting question how such sensitive anisotropy dependence in the classical model can reflect on the corresponding quantum spin system with various XY anisotropy.

Although from a view point of the spin wave theory, quantum fluctuations should cause an effect 
similar to that of thermal fluctuations\cite{Chubukov}, it is still interesting to see 
how it appears in strongly fluctuating quantum systems.
In this paper, we study 
whether similar phase structure with strong dependence on the anisotropy appears or not 
by investigation of the ground state properties in the magnetic field 
by various numerical methods.
In the case of three-leg ladder system, which well represents the triangular lattice, 
the dependence is certainly similar to those found in the corresponding classical systems at finite temperatures as the entropy effect.
We also study the dependence of the ground state on the lattice shape.
In the classical system, the ground state configuration of all the lattices consisting of local triangles is an assembly of local ground state configuration of three spins representing the three sublattices of the triangular lattice. 
Thus, the ground state configurations of those lattices 
have the same spin configuration.
By contrast,  the interaction in quantum systems is not
local because of the non-commutativity among the spins, and
the ground state configuration depends on the lattice shape
even if they locally consist of the same triangles.
In fact, we demonstrate such dependence in a two-leg ladder system.

In order to study the quantum ground state of the present model, we cannot apply the quantum Monte Carlo simulation because of the negative-sign problem.
Therefore, in this paper, we adopt kinds of diagonalization method, i.e., exact diagonalization method for a finite lattice of small number of spins, the product wave-function renormalization group (PWFRG) method which is a variant of density matrix renormalization group (DMRG) method.\cite{PWFRG,DMRG} PWFRG has been developed to study the magnetization process in the ground state. \cite{PWFRG} We also use the DMRG method for finite lattices which gives complementary data for those from PWFRG.

This paper is organized as follows. We explain the model studied in the present paper 
and order parameters
in \S 2, where corresponding data in the classical model are also reviewed.
In \S 3, the dependence of the ground state configuration on the anisotropy
and on the external field is studied in a finite lattice.
In \S 4, the dependence is studied in a two-leg ladder system which
is an assembly of triangles.
In \S 5, properties on a three-leg ladder 
with the same periodicity as the triangular lattice is studied.
In \S 6, summary and discussion are given.    

\section{Model and Order parameters}

For the study of CsCuCl$_3$, three-dimensional XY-like Heisenberg model
has been adopted. 
In the classical model,
the anisotropy is usually introduced by the single-ion anisotropy
\begin{equation}
{\cal H}=\sum_{\langle ij\rangle}J_{ij}{\mbold S}_i\cdot {\mbold S}_j
-D\sum_i(S_i^z)^2-H\sum_iS_i^z,
\label{model}
\end{equation}
where $\langle ij\rangle$ denotes the interaction in a triangular layer.
The interaction in the layer is antiferromagnetic, i.e.,
$J_{ij}=J>0$,
while the  interaction for the interlayer interaction is ferromagnetic, i.e., 
$J_{ij}=J'<0$. For simplicity we take  $J'=-J$.

In the case of $S=1/2$ quantum system, however, the single-ion anisotropy 
does not exist. Therefore, we adopt anisotropic coupling as a source of the anisotropy.
In this paper, we study only two-dimensional systems because of the 
limit of the capacity of numerical calculation. The Hamiltonian is given by
\begin{equation}
{\cal H}=\sum_{\langle ij\rangle}J_{ij}\left(S_i^xS_j^x+S_i^yS_j^y+AS_i^zS_j^z\right)
-H\sum_iS_i^z,
\label{model2D}
\end{equation}
where $S_i^{\alpha}={1\over2}\sigma_i^{\alpha}, (\alpha=x,y$ and $z)$ where
$\sigma_i^{\alpha}$ is the $\alpha$ component of the Pauli matrix.
The magnetization of the system is measured by
\begin{equation}
M=\sum_iS_i^z.
\end{equation}

\begin{figure}
$$
\epsfxsize=8cm \epsfysize=3.0cm \epsfbox{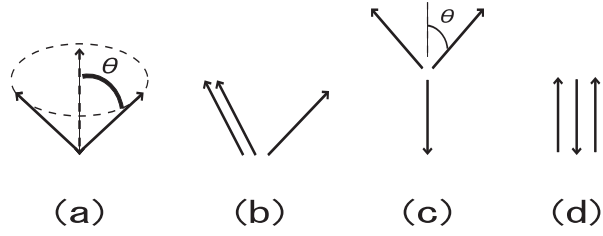}$$
\vspace {-0.5cm}
\caption{Candidates for the ground state configurations:
(a) umbrella structure, 
(b) v-shape structure.
(c) Y-shape structure, and
(d) collinear ferrimagnetic structure.
}
\label{GSconf}
\end{figure}

Low-energy configurations of the model (\ref{GSconf}) in the classical 
case is depicted in Figs.\ref{GSconf}, where the angles $\theta$ are respectively, given by 
\begin{eqnarray}
\cos\theta& ={H\over 3(2A+1)} &\quad {\rm for\ Fig.~\ref{GSconf}(a)},\\
\cos\theta&\!\!={3A+H\over 3(A+1)}& \quad {\rm for\ Fig.~\ref{GSconf}(c)}.
\end{eqnarray}
For the configuration Fig.~\ref{GSconf}(b), we have to calculate $\theta$ numerically.
Comparing energies of the configurations, 
we find that the umbrella configuration (Fig.~\ref{GSconf}(a))
gives the minimum energy for $A<1$, corresponding to $D<0$ in the classical spin model (\ref{model}).
\vspace*{.5cm}

In order to characterize the phases, we define
the following order parameters.
The unit cell in the ordered phases consists of three sublattices.
First, we introduce the chirality\cite {Miyashita-Shiba}:
\begin{equation}
\mbold {\kappa}={2\over 3\sqrt{2}}\sum_{i\in{\rm A},j\in{\rm B},k\in{\rm C}}
(\mbold S_{i}\times \mbold S_{j}+
\mbold S_{j}\times \mbold S_{k}+\mbold S_{k}\times \mbold S_{i}),
\end{equation}
where $i,j$ and $k$ denote sites on a triangle consisting of the sublattices A, B, and C
of the triangular lattice, respectively.
The presence of the umbrella structure can be identified by measuring
the $z$ component of the chirality.

In the v-shape structure, the spins on the two of sublattices are parallel, and
thus the chirality disappears.
In order to identify noncollinear ordering structures of the v-shape structure,
we measure the following quantities:
\begin{equation}\label{eq:finite}
\mbold{X}^{\rm AB}=\sum_{i\in {\rm A},j\in {\rm B}}\mbold{S}_i\times \mbold{S}_j,\quad
\mbold{X}^{\rm BC}=\sum_{i\in {\rm B},j\in {\rm C}}\mbold{S}_i\times \mbold{S}_j,\quad
{\rm and }\quad
\mbold{X}^{\rm CA}=\sum_{i\in {\rm C},j\in {\rm A}}\mbold{S}_i\times \mbold{S}_j.
\end{equation}
We define the following order parameter to detect the non-collinearity
\begin{equation}
X_{\alpha}\equiv {1\over N^2}
\left[ ({X}^{\rm AB}_{\alpha})^2+({X}^{\rm BC}_{\alpha})^2+({X}^{\rm
CA}_{\alpha})^2\right],
\end{equation}
where $\alpha=x, y, {\rm and } \ z$, and $N$ is the number of spins.
In particular, we use
an order parameter
\begin{equation}
S_{xy}=X_x+X_y
\label{Sxy}
\end{equation}
to detect the v-shape structure when the field is applied in the $z$ direction.

If the configuration is of the umbrella type, the order parameters
\begin{equation}
K_z\equiv {1\over N^2}\langle \kappa_z^2\rangle
\label{kappaz}
\end{equation}
and $S_{xy}$ are nonzero in the limit $N\rightarrow \infty$.
On the other hand, in the v-shape structure, $K_{z}$ is zero while
$S_{xy}$ remains non-zero. 

The $xy$ component of the chirality $\mbold{\kappa}$ is good 
to detect the Y-shape structure. Therefore, 
we also observe the following quantity: 
\begin{equation}
K_{xy}\equiv {1\over N^2} (\langle \kappa_x^2\rangle + \langle \kappa_y^2\rangle).
\label{kappaxy}
\end{equation}
This order parameter is non-zero in the Y-shape structure, but it is
zero in the v-shape structure.

\begin{figure}
$$\begin{array}{cc}
\epsfxsize=6cm \epsfysize=5.0cm \epsfbox{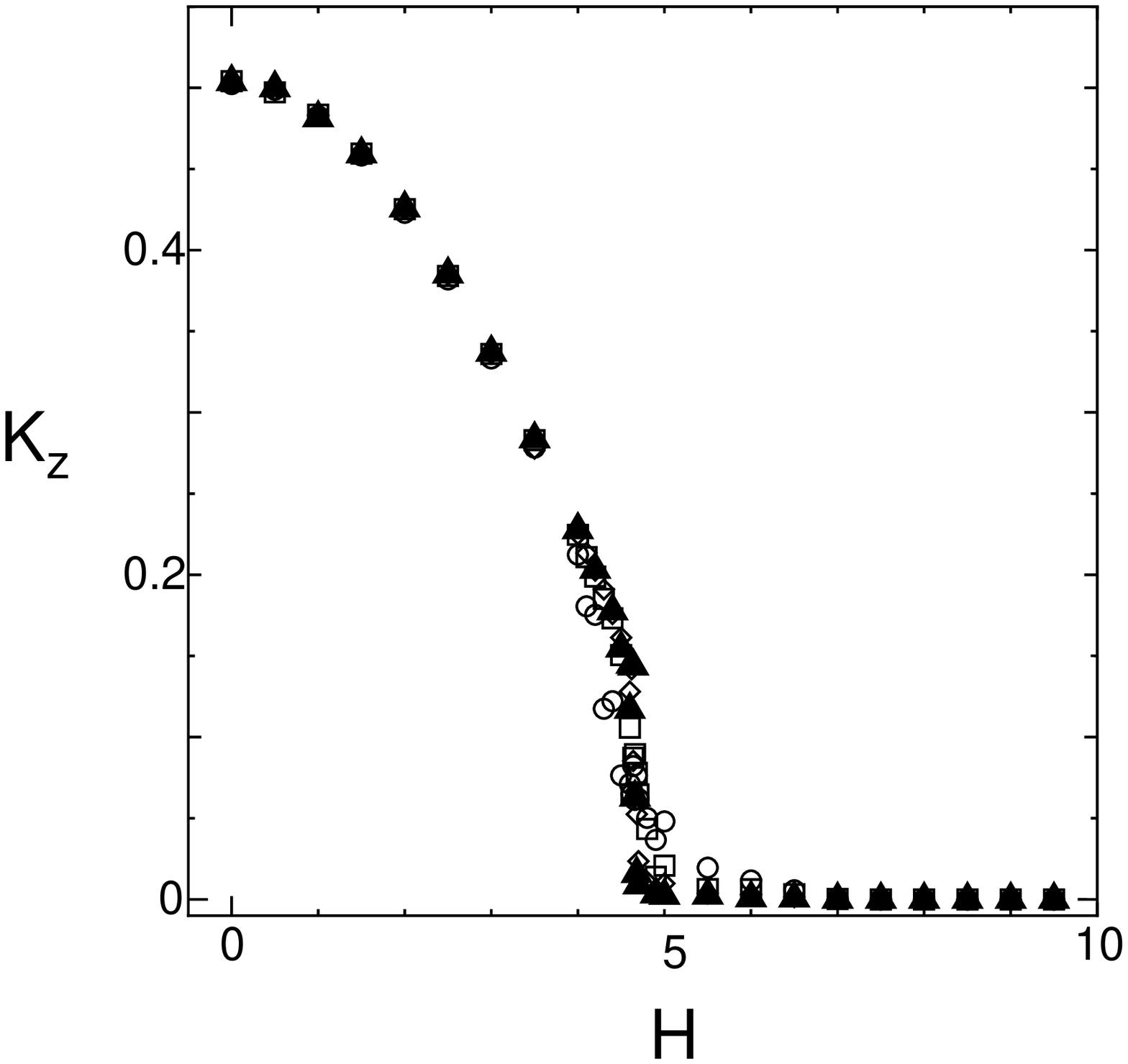} &
\epsfxsize=6cm \epsfysize=5.0cm \epsfbox{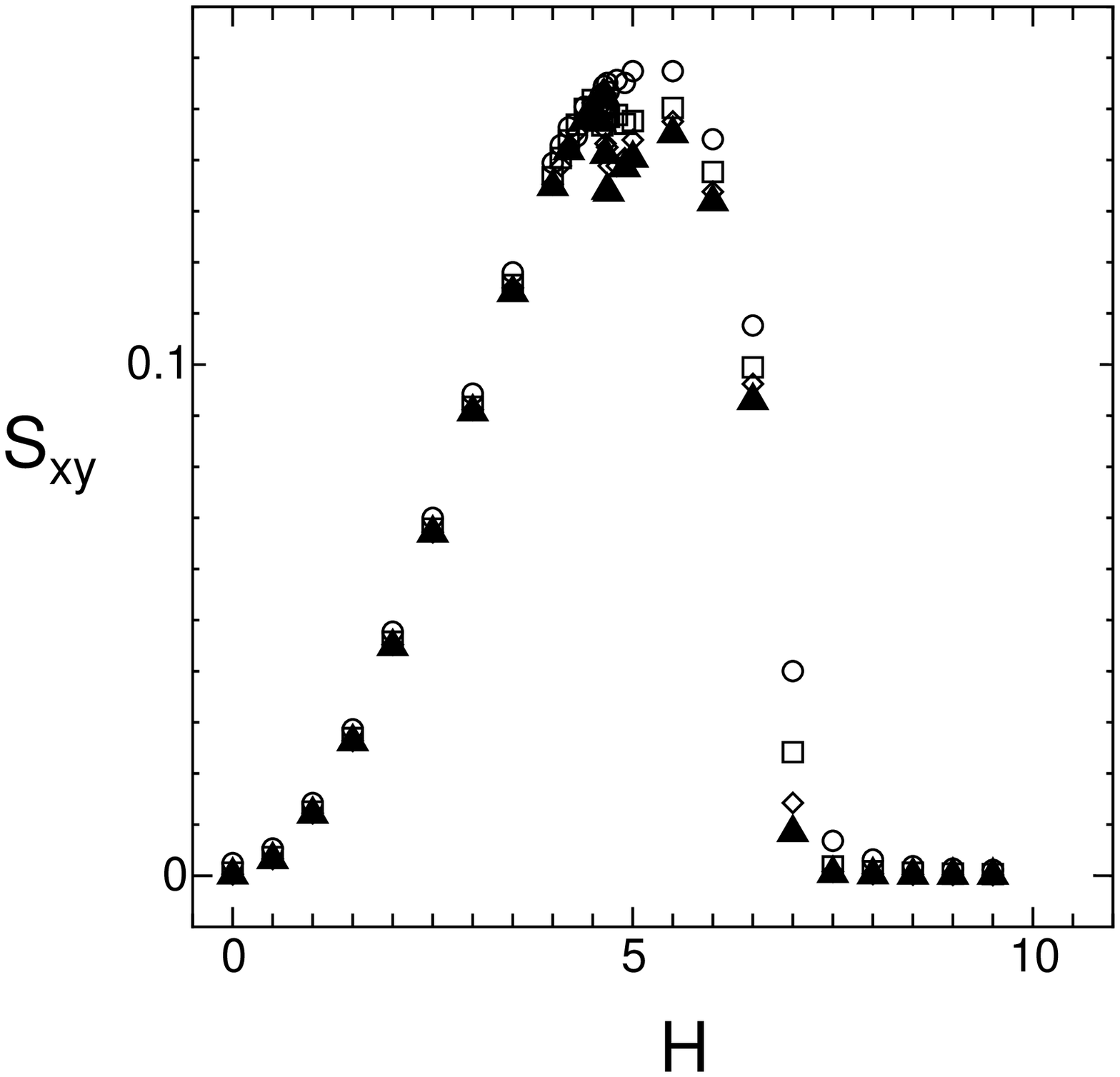}\\
{\rm (a)} & {\rm (b)}\\
& \\ 
\epsfxsize=6cm \epsfysize=5.0cm \epsfbox{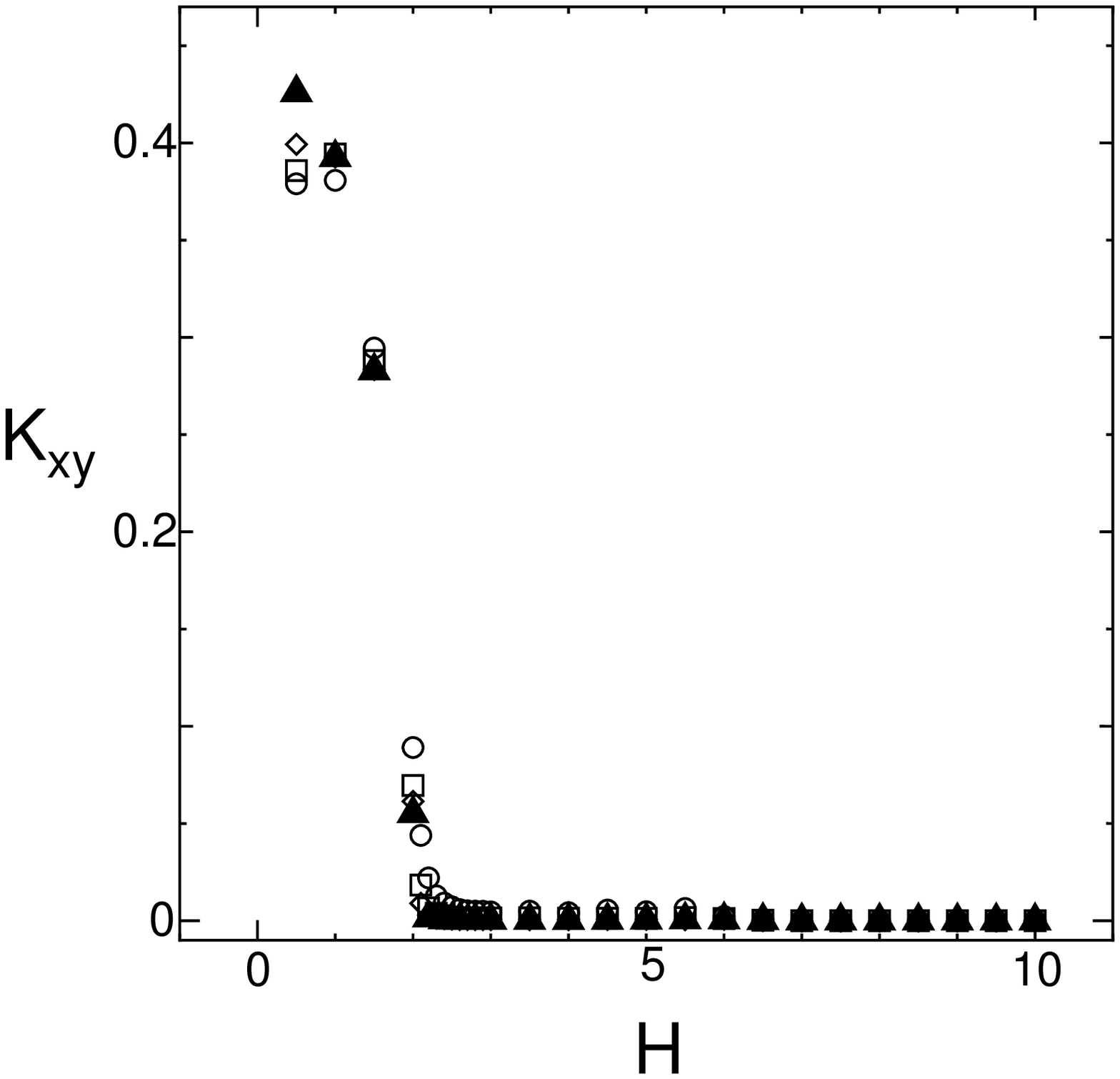} &
\epsfxsize=6cm \epsfysize=5.0cm \epsfbox{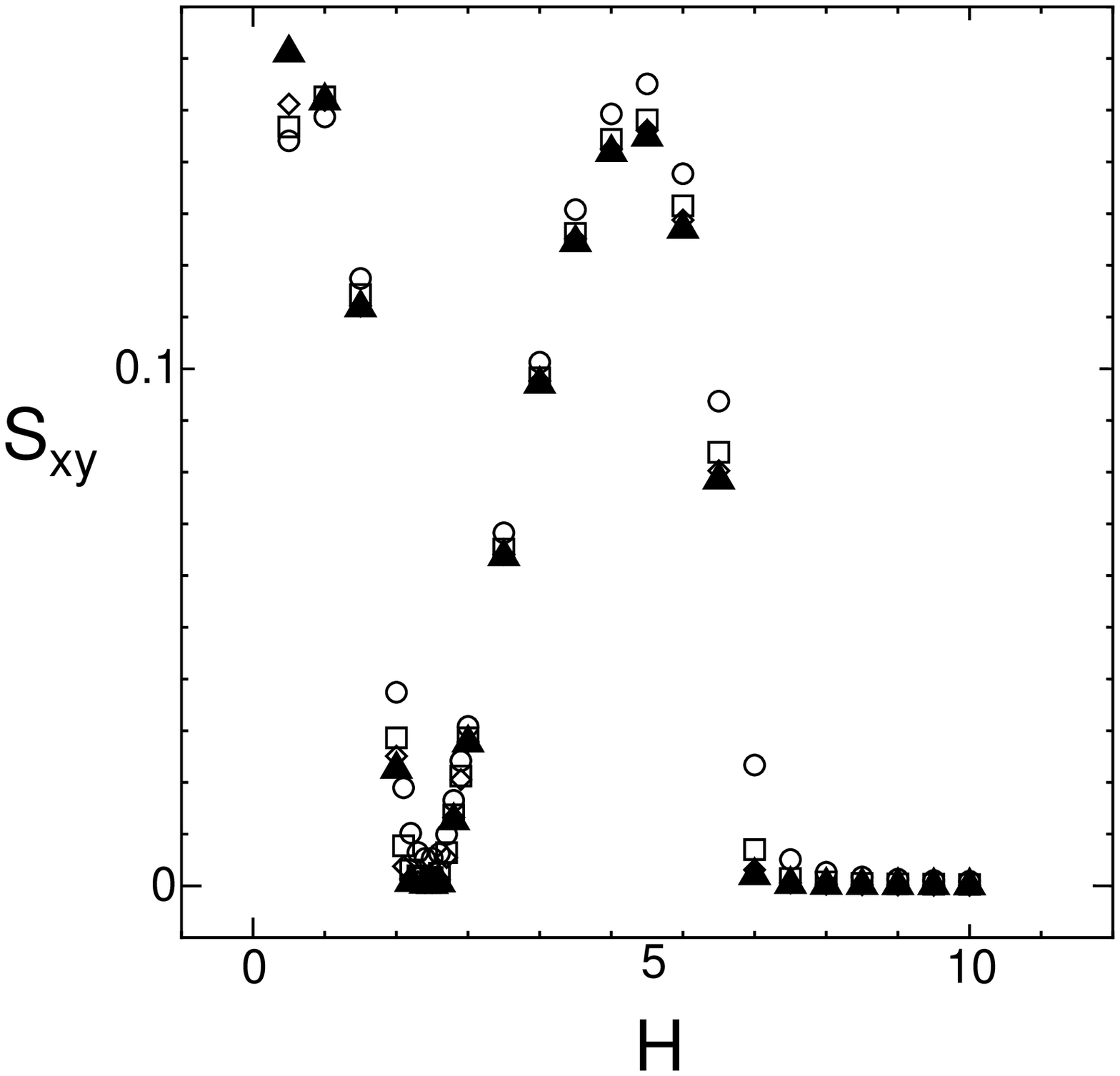}\\
{\rm (c)} & {\rm (d)} 
\end{array} $$
\caption{The field dependence of (a) $\kappa_z$ and
(b) $S_{xy}$ at $T=0.5$ for the weakly XY-anisotropic case ($D=-0.1$).
(c) $\kappa_{xy}$ and
(d) $S_{xy}$ at $T=0.5$ for the Heisenberg model ($D=0$).
The symbols, circle, square,diamond, and (closed) triangle
denote the size $L=$ 12, 18, 24, and 30, respectively.
(from the reference \cite{WMS})
}
\label{XYD01}
\end{figure}

For the classical model (\ref {model}), we can perform Monte Carlo simulations
in large systems and determine the thermodynamic stable state.
In Figs.~\ref {XYD01}, the field-dependence of the chirality and $S_{xy}$ of a weakly XY-like Heisenberg model, i.e., 
$D/J=-0.1$, at temperature $T/J$=0.5, are depicted. 
There, as the field increases, 
the structure changes
from the umbrella to the v-shape discontinuously and finally to the field-induced ferromagnetic state.
The data for the Heisenberg model are also depicted 
in Figs.~\ref{XYD01}(c) and (d).\cite{WMS}
There the structure changes from the Y-shape, collinear up-up-down,
and the v-shape, and finally to the field-induced ferromagnetic state. 

\vspace*{1cm}

\section{Observation on a finite lattice}\label{ch:finite}

\begin{figure}
$$\begin{array}{cc}
\epsfxsize=6cm \epsfysize=4.0cm \epsfbox{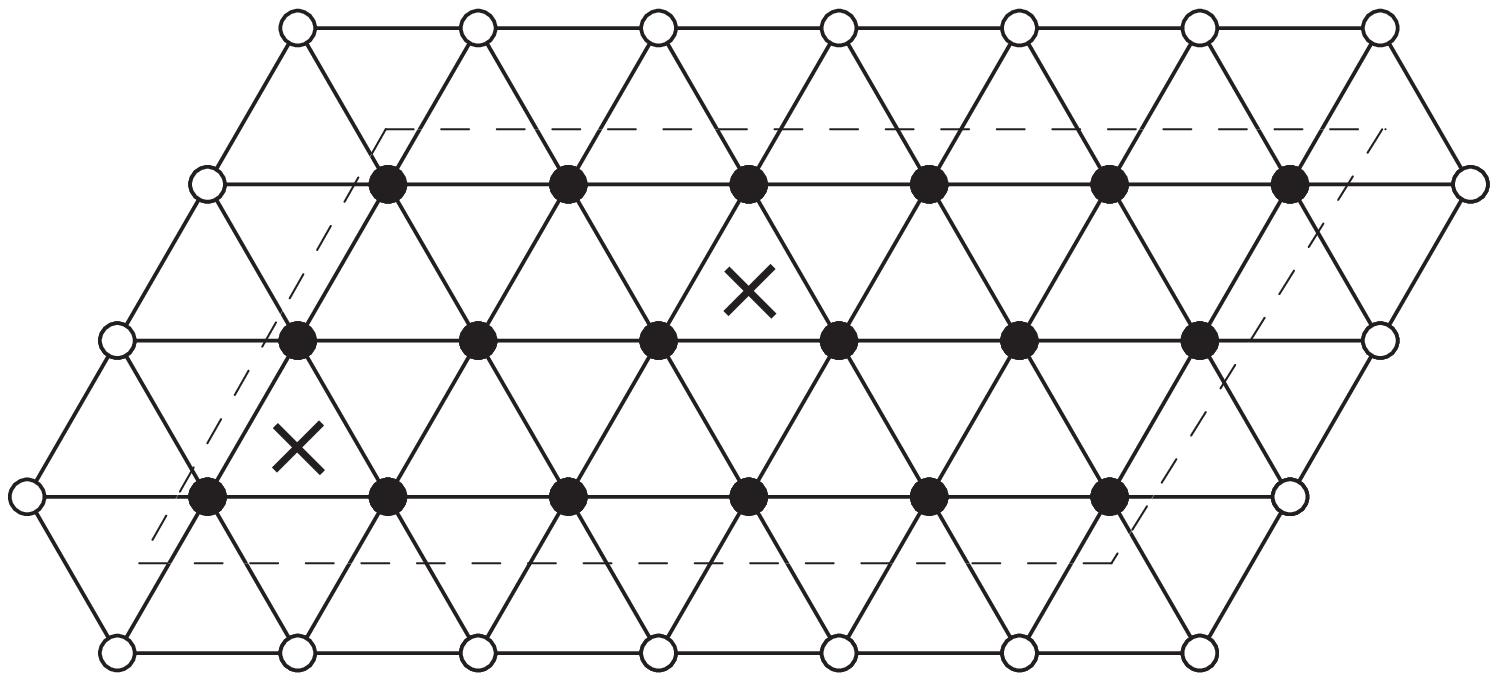}&
\epsfxsize=6cm \epsfysize=5.0cm \epsfbox{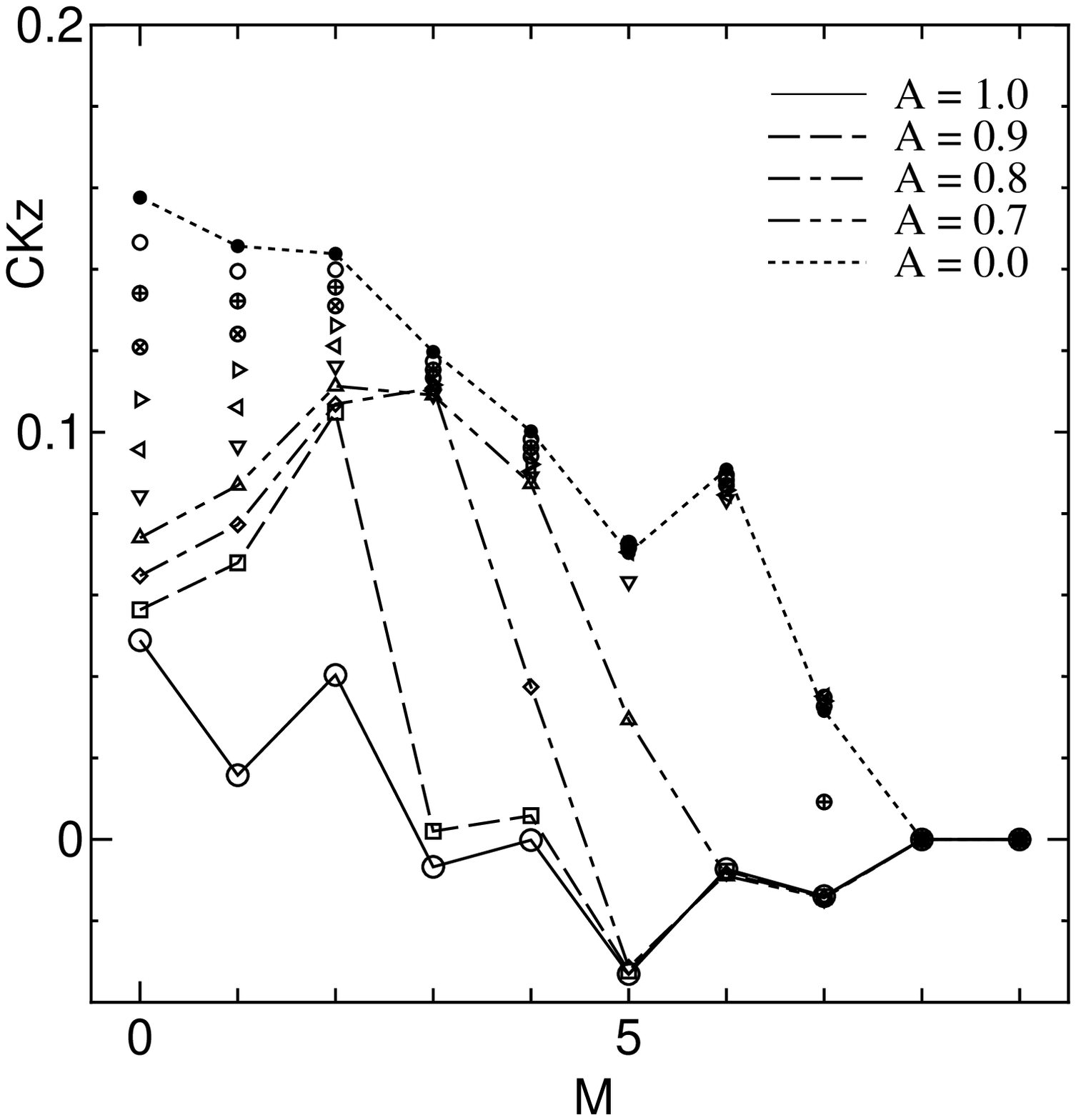} \\
{\rm (a)} & {\rm (b)} \\
\epsfxsize=6cm \epsfysize=5.0cm \epsfbox{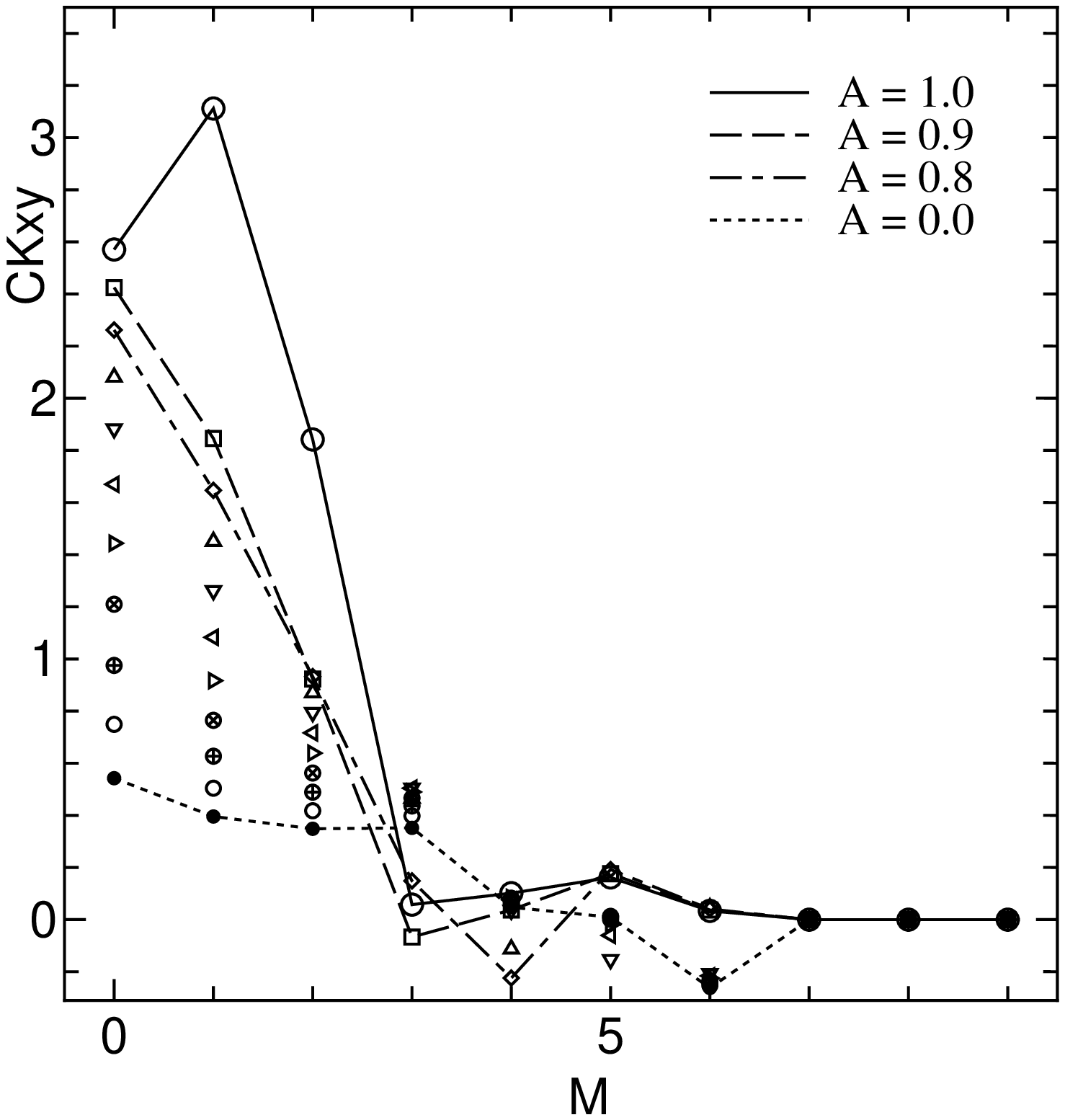}&
\epsfxsize=6cm \epsfysize=5.0cm \epsfbox{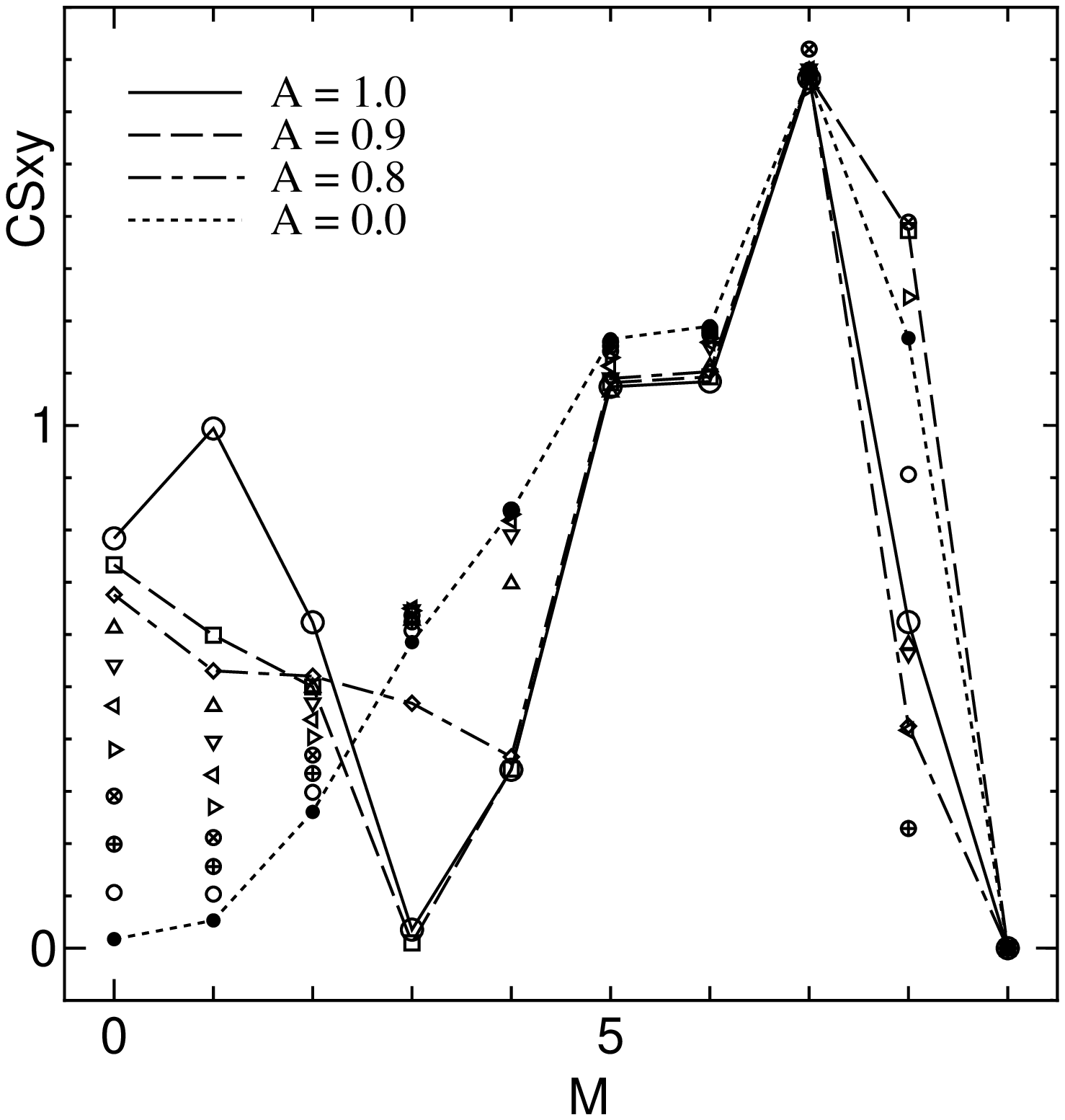} \\
{\rm (c)} &  {\rm (d)} 
\end{array} $$
\vspace{0.3cm}
\label{fig-Data18}
\caption{
(a)The finite lattice for the diagonalization study.
The correlation functions of the order parameters are obtained at 
the quantities on the positions denoted by crosses.
The field dependence of (b) $CK_{z}$ and
(c) $CK_{xy}$, and (d) $CS_{xy}$ as functions of the
magnetization $M$. Symbols denote the data for various values of $A$:
{$\bullet $}(A=0.0),{$\circ $}(0.1),{$\oplus $}(0.2),{$\otimes $}(0.3),
{$\rhd $}(0.4),{$\lhd $}(0.5),{$\bigtriangledown $}(0.6),
{$\bigtriangleup $}(0.7),
{$\Diamond $}(0.8),{$\Box $}(0.9), and $\bigcirc $(1.0)}
\end{figure}

For the present quantum systems, Monte Carlo simulation
is not available because of the negative-sign problem.
Thus, the most naive approach to study the ground state is 
a diagonalization of the Hamiltonian of a finite lattice of small number of spins.
In this section, we study systems in a lattice of $3\times 6$ with
the periodic boundary condition depicted in Fig.~\ref{fig-Data18}(a).
For the finite size lattice, we study the property of the state with a fixed 
magnetization $M$ instead of the magnetic field $H$.
The change of $M$ corresponds to that of $H$ because the magnetization
of the ground state monotonically increases with the field.
We note that, in Refs.\cite{honecker,cabra}, magnetization processes of the similar systems have been also studied for finite size and/or ladder lattices.

In a finite lattice, we investigate correlation functions 
of the quantities corresponding to the order parameters, 
because the expectation values of the order parameters themselves
may vanish in the quantum ground state.
Here we observe the correlation functions of the $z$-component of the chirality 
($CK_{z}$), the $xy$-component of the chirality ($CK_{xy}$), and 
the $xy$-component of the vector product of spins defined in (\ref{eq:finite}) 
($CS_{xy}$), 
at the sites denoted in Fig.~\ref{fig-Data18}(a). $CK_{z},CK_{xy}$, and $CS_{xy}$ 
are given by
\begin{eqnarray}
CK_{z}\!\!\!\!\!\! 
&\equiv & 
\!\!\!\!\!\!
\small{
\left\langle
\left[ 
\textbf{S}_1\times \textbf{S}_2 +  
\textbf{S}_2\times \textbf{S}_3 + 
\textbf{S}_3\times \textbf{S}_1 
\right]_z 
\cdot  
\left[ 
\textbf{S}_{1^{'}}\times \textbf{S}_{2^{'}} +  
\textbf{S}_{2^{'}}\times \textbf{S}_{3^{'}} + 
\textbf{S}_{3^{'}}\times \textbf{S}_{1^{'}} 
\right]_z 
\right\rangle 
}
\\
CK_{xy}\!\!\!\!\!\! 
&\equiv& 
\!\!\!\!\!\!
\small{
\left\langle
\left[ 
\textbf{S}_1\times \textbf{S}_2 +  
\textbf{S}_2\times \textbf{S}_3 + 
\textbf{S}_3\times \textbf{S}_1 
\right]_x 
\cdot 
\left[ 
\textbf{S}_{1^{'}}\times \textbf{S}_{2^{'}} +  
\textbf{S}_{2^{'}}\times \textbf{S}_{3^{'}} + 
\textbf{S}_{3^{'}}\times \textbf{S}_{1^{'}} 
\right]_x 
\right\rangle 
}
\nonumber \\ 
&+&
\!\!\!\!\!\!
\small{
\left\langle
\left[ 
\textbf{S}_1\times \textbf{S}_2 +  
\textbf{S}_2\times \textbf{S}_3 + 
\textbf{S}_3\times \textbf{S}_1 
\right]_y 
\cdot 
\left[ 
\textbf{S}_{1^{'}}\times \textbf{S}_{2^{'}} +  
\textbf{S}_{2^{'}}\times \textbf{S}_{3^{'}} + 
\textbf{S}_{3^{'}}\times \textbf{S}_{1^{'}} 
\right]_y 
\right\rangle 
} \\
CS_{xy}\!\!\!\!\!\! 
&\equiv & 
\!\!\!\!\!\!
\small{
\left\langle
\left[ 
\textbf{S}_1\times \textbf{S}_2   
\right]_x 
\!\!
\cdot 
\!\!
\left[ 
\textbf{S}_{1^{'}}\times \textbf{S}_{2^{'}} 
\right]_x 
\!\!
+
\left[ 
\textbf{S}_2\times \textbf{S}_3   
\right]_x 
\!\!
\cdot 
\!\!
\left[ 
\textbf{S}_{2^{'}}\times \textbf{S}_{3^{'}} 
\right]_x 
\!\!
+
\left[ 
\textbf{S}_3\times \textbf{S}_1   
\right]_x 
\!\!
\cdot 
\!\! 
\left[ 
\textbf{S}_{3^{'}}\times \textbf{S}_{1^{'}} 
\right]_x 
\right\rangle 
}
\nonumber \\ 
&+&
\!\!\!\!\!\!
\small{
\left\langle
\left[ 
\textbf{S}_1\times \textbf{S}_2 
\right]_y 
\!\!
\cdot 
\!\!
\left[ 
\textbf{S}_{1^{'}}\times \textbf{S}_{2^{'}} 
\right]_y 
\!\!
+
\left[ 
\textbf{S}_2\times \textbf{S}_3 
\right]_y 
\!\!
\cdot 
\!\!
\left[ 
\textbf{S}_{2^{'}}\times \textbf{S}_{3^{'}} 
\right]_y 
\!\!
+
\left[ 
\textbf{S}_3\times \textbf{S}_1 
\right]_y 
\!\!
\cdot 
\!\!
\left[ 
\textbf{S}_{3^{'}}\times \textbf{S}_{1^{'}} 
\right]_y 
\right\rangle 
},
\end{eqnarray}
where 1,2,3 and 1$^{'}$,2$^{'}$,3$^{'}$ are the lattice sites of the triangles 
denoted by crosses.
We plot  $CK_{z},CK_{xy}$ and $CS_{xy}$ as functions of the magnetization for various values of the anisotropic coupling $A$ $(A=0,0.1,\cdots,1)$  in Fig.~\ref{fig-Data18}(b), (c), and (d), respectively.

We find a qualitative similar dependence of the order parameters to those 
in the classical models.
That is, $CK_z$ decreases with the magnetization $M$ smoothly for 
the case of strongly anisotropy $(A=0)$, where no transition occurs until the
saturation field $H_{\rm c0}$. 

When the system has weak anisotropy, e.g. $A=0.9$, then 
$CK_z$ decreases suddenly at an intermediate value of
the magnetization $M$, although $CS_{xy}$ is still large.
This case corresponds to the classical case depicted 
in Figs.~\ref{XYD01}(a) and Figs.~\ref{XYD01}(b).
At the Heisenberg case ($A=1$), instead of $CK_{z}$,
$CK_{xy}$ appears at small magnetizations, and it disappears at
an intermediate field where the collinear up-up-down state appears.
For larger $M$, another non-collinear state appears.
These dependences agree with those in Figs.~\ref {XYD01}(c) and 
Figs.~\ref {XYD01}(d).  
Thus, the anisotropy and field dependence of the ground state property is similar to that in the classical systems, although number of data points are limited because only the finite lattice is studied. 
In the next section, we study systems consisting of infinite number of triangles.  

\section{Observation on a ladder system consisting of triangles}

\begin{figure}
$$
\epsfxsize=8cm \epsfysize=3.0cm \epsfbox{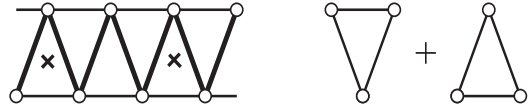}$$

\caption{Two-leg ladder consisting of triangles.
The correlation functions of the order parameters are obtained at 
the quantities on the positions denoted by crosses.}
\label{fig-Lattice-ladder}
\end{figure}

In this section, we study a two-leg ladder system with zigzag structure, which consists of triangles depicted in Fig.~\ref{fig-Lattice-ladder}.
The Hamiltonian of this case is given by
\begin{equation}
{\cal H}=\sum_{i}\left(
2J(S_i^xS_{i+1}^x+S_i^yS_{i+1}^y+AS_i^zS_{i+1}^z)+
J(S_i^xS_{i+2}^x+S_i^yS_{i+2}^y+AS_i^zS_{i+2}^z)\right)
-H\sum_iS_i^z.
\end{equation}
Heisenberg antiferromagnets with $A=1$ 
in this type of lattice have been 
studied in detail as the  Majumdar-Ghosh model.\cite{MD}
There, various interesting properties, e.g.,
energy gap between the ground state and the first excited state, and cusp singularities in the magnetic process, have been found.\cite{zigzag} 
Here, we study properties of the XY-like antiferromagnets on this lattice ($A<1$).
For this type of one dimensional lattices, methods of the DMRG type
are very efficient. Here, we study the magnetization process by
PWFRG and DMRG.
As in the exact diagonalization for a finite size lattice, the expectation values of the order parameters themselves may vanish in the quantum case. 
We, therefore, measure the correlation functions $CK_z$ and $CS_{xy}$
between two triangles near the center of the ladder, which are indicated by $\times$ symbols in Fig.\ref{fig-Lattice-ladder}.

\begin{figure}
$$\begin{array}{ccc}
\epsfxsize=5cm \epsfysize=4.5cm \epsfbox{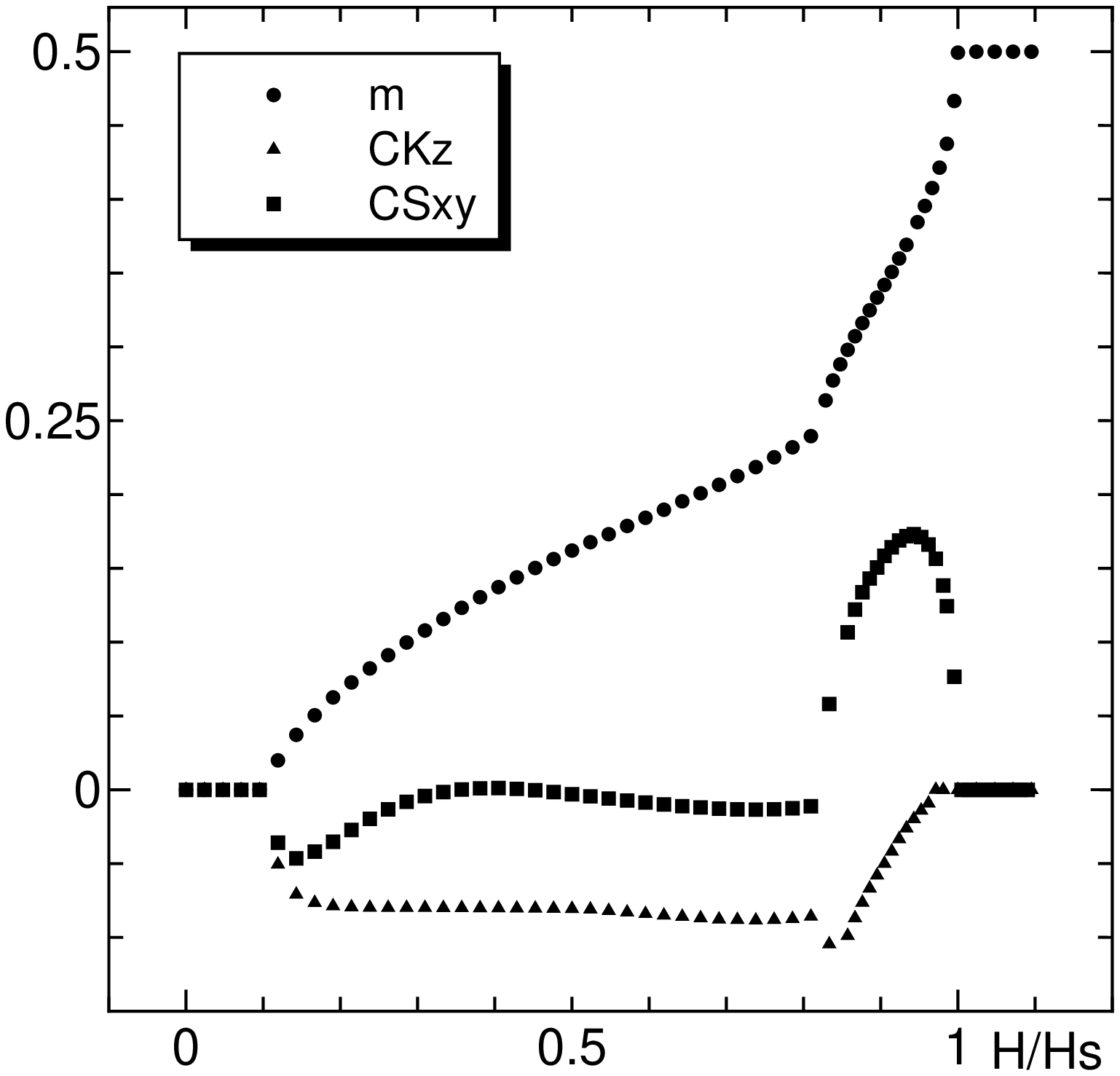} &
\epsfxsize=5cm \epsfysize=4.5cm \epsfbox{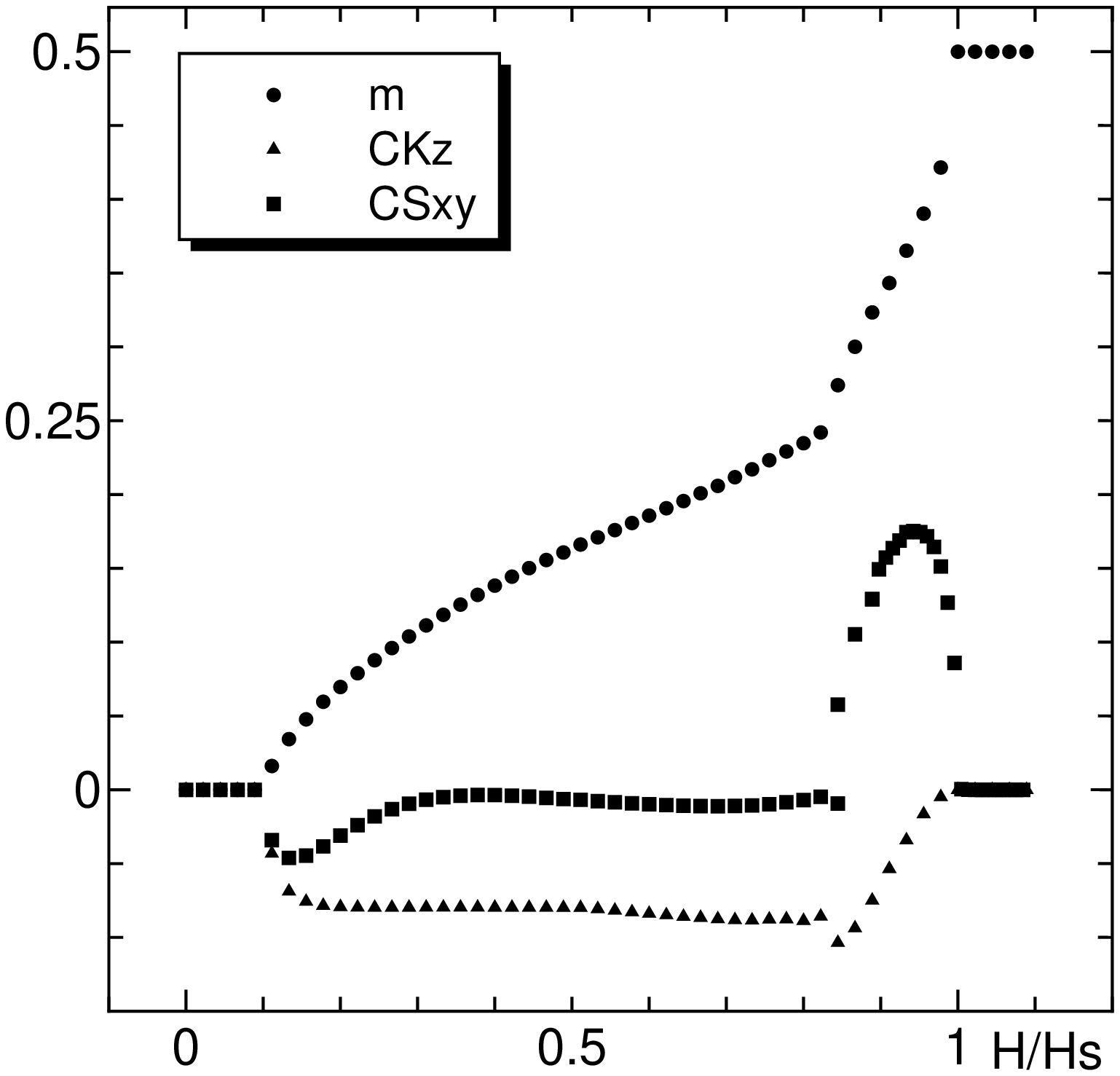} &
\epsfxsize=5cm \epsfysize=4.5cm \epsfbox{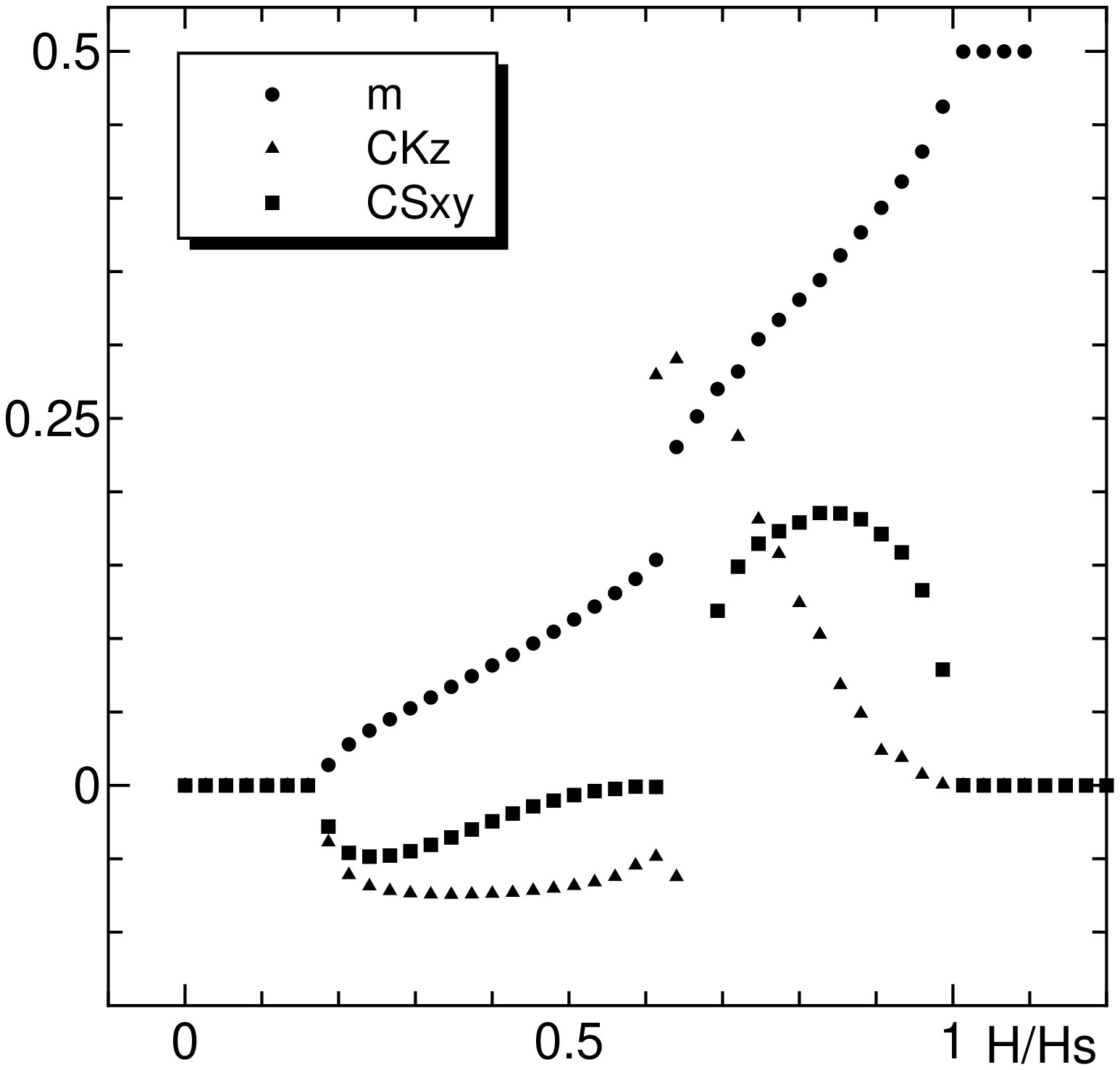}\\
{\rm (a)} & {\rm (b)} & {\rm (c)} 
\end{array} $$
\caption{The field dependence of the magnetization $m$, $CK_z$ and $CS_{xy}$ as functions of the field  $H$ for (a) $A=0.9$, (b) $A=1.0$ and (c) $A=0.0$}
\label{fig-ladder2XY}
\end{figure}

In Fig.~\ref{fig-ladder2XY}(a), we depict the magnetization for $A=0.9$
obtained by PWFRG. In the present study we use a block Hamiltonian of the dimension $m=48$. We have confirmed that the results converge up to  this value of $m$. 
In the magnetization process, we find a zero magnetization region at low field, which implies the existence of energy gap.
This gap is due to the dimering effect that brings the Majumdar-Ghosh solution in the Heisenberg limit.\cite{MD}
In Fig.~\ref{fig-ladder2XY}(a), 
the correlation functions of the order parameters $CK_{z}$ and $CS_{xy}$ are also plotted.
They behave very differently from those expected in the classical lattice. 
This may be due to the fact that the zigzag chain does not have the three sublattice structure in the rung direction.
For example $CK_z$ takes a negative value for $A>0$, for which there is no classical analogue.
Thus, in the next section we study a lattice which is more close to
the triangular lattice.

\begin{figure}
$$\begin{array}{ccc}
\epsfxsize=5cm \epsfysize=4.5cm \epsfbox{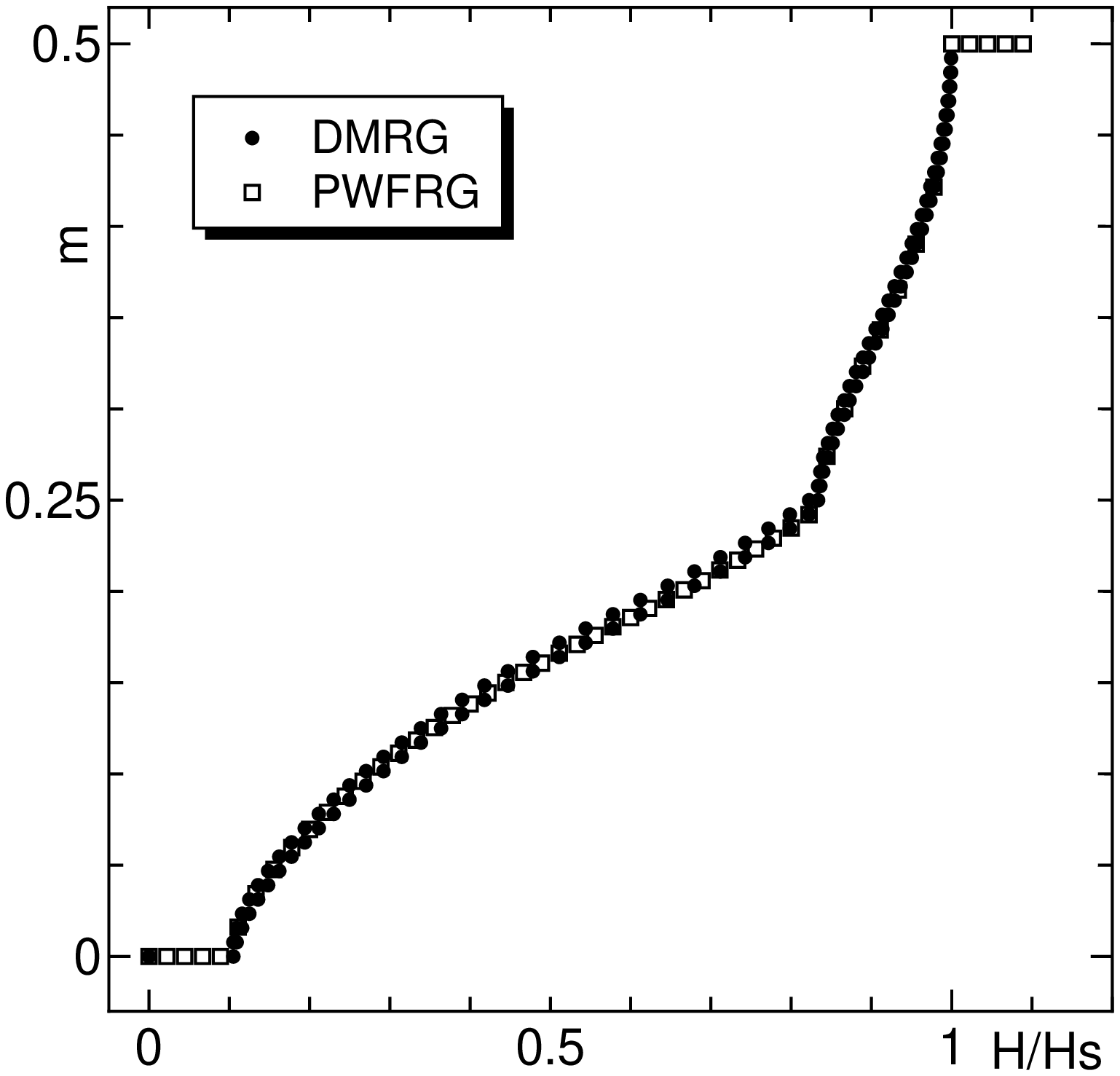} &
\epsfxsize=5cm \epsfysize=4.5cm \epsfbox{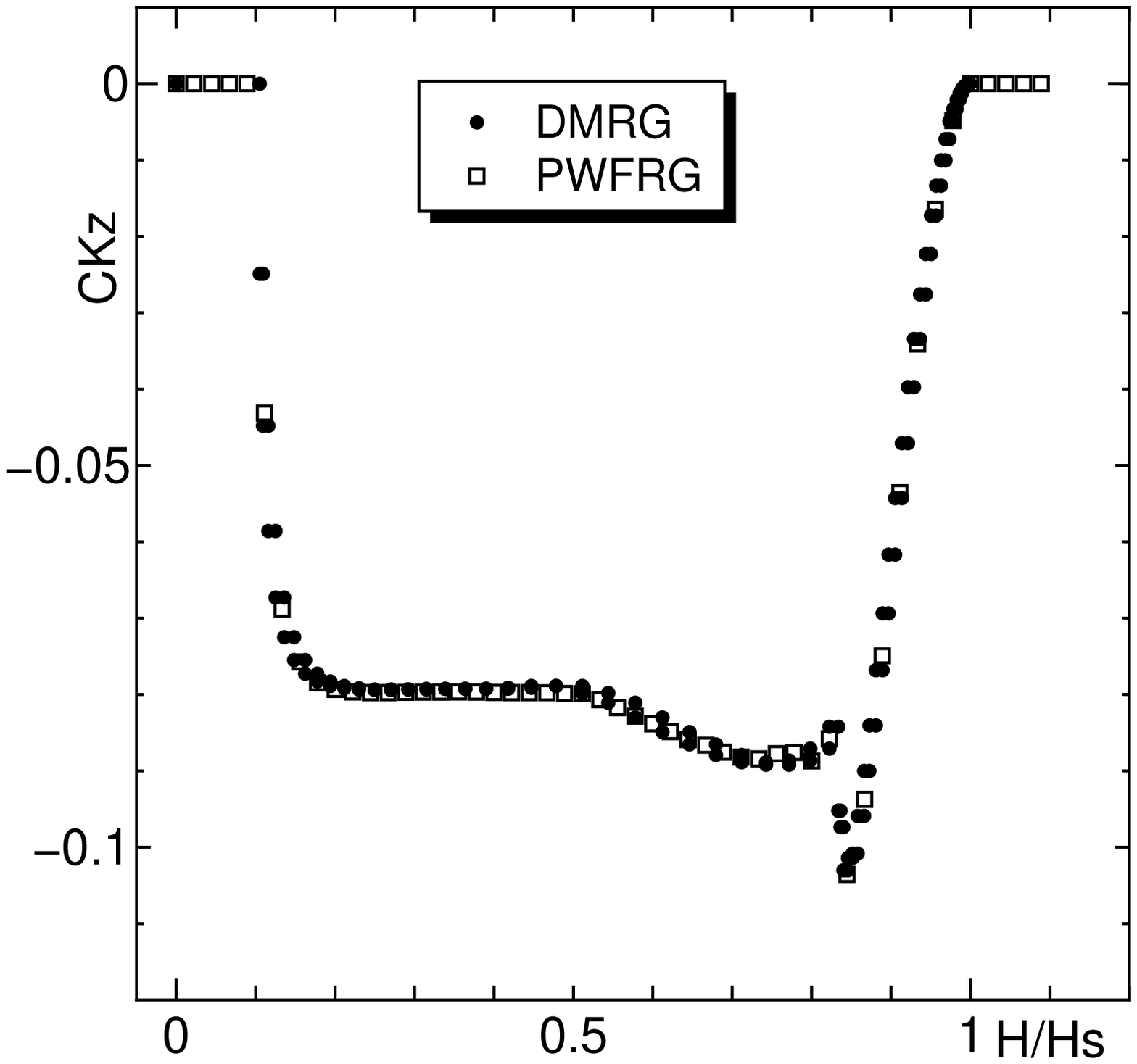} &
\epsfxsize=5cm \epsfysize=4.5cm \epsfbox{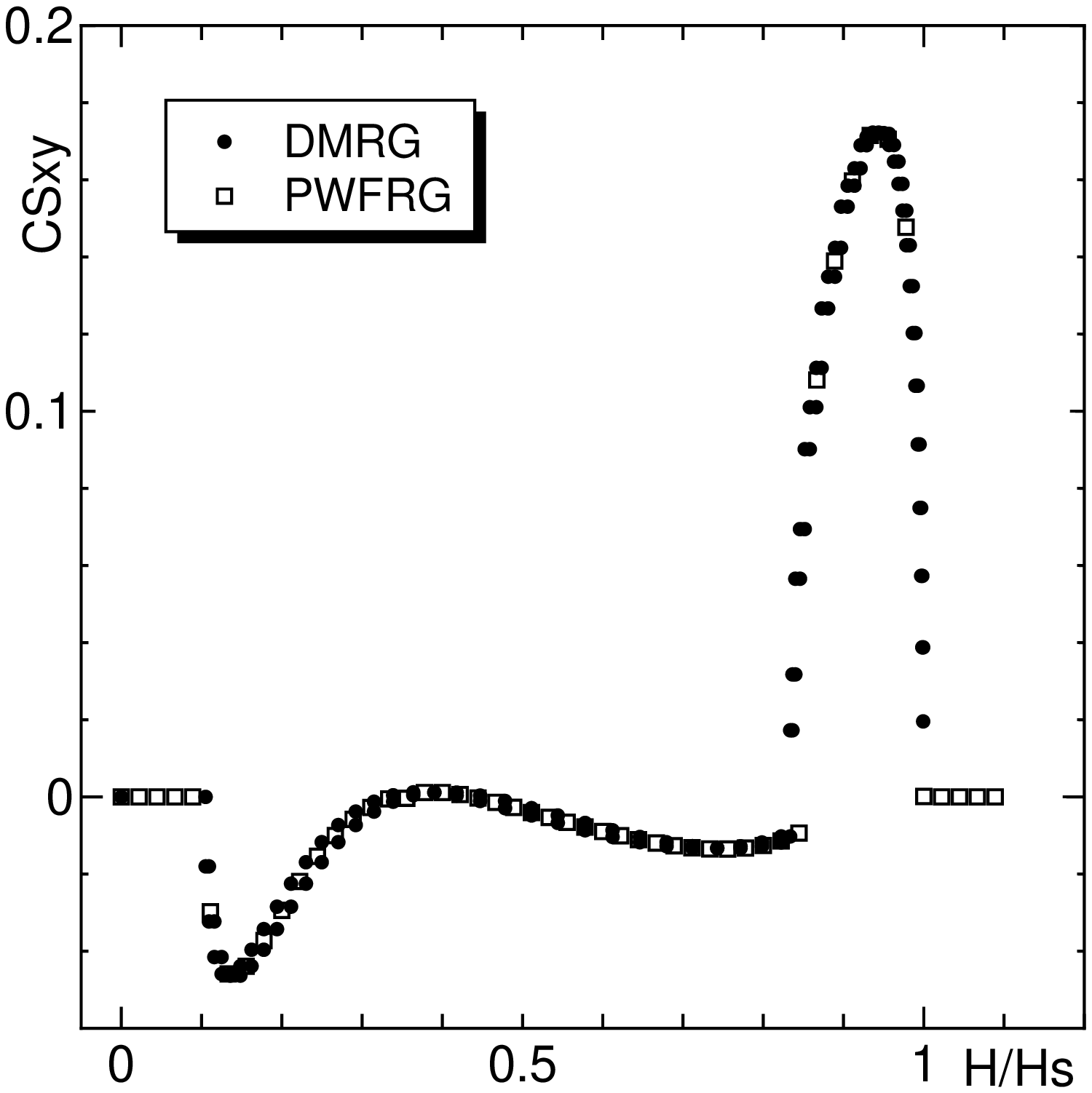}\\
{\rm (a)} & {\rm (b)} & {\rm (c)} 
\end{array} $$
\caption{The field dependence of (a) the magnetization $m$, (b) $CK_z$ and
(c) $CS_{xy}$ as functions of the field at $A=1$ obtained by PWFRG and DMRG.}
\label{fig-PWFRG-DMRG}
\end{figure}

Before proceeding to the next section, here we make a comparison between the results for PWFRG and DMRG. 
In PWFRG, we obtain the magnitude of the magnetization for the given field. 
On the other hand, in the DMRG procedure,\cite{DMRG} we
obtain a ground state energy for given values of the magnetization
from which we can obtain the magnetization curve as a function of
the field because the magnetization commutes with the Hamiltonian
for the present model. Both methods should give the same results.
In Figs.~\ref {fig-PWFRG-DMRG}, we demonstrate that the PWFRG and DMRG results  agree with each other by showing the quantities for the Majumdar-Ghosh model of $L=128$ system obtained by the DMRG with $m=64$.
In the figure   we take spatial averages of $CK_z$ and $CS_{xy}$ for the comparison with the PWFRG results, since $CK_z$ and $CS_{xy}$ in a finite size system show position depending behavior due to the open boundary. 

\section{Three-leg triangular lattice}

\begin{figure}
$$
\epsfxsize=7cm \epsfysize=4cm \epsfbox{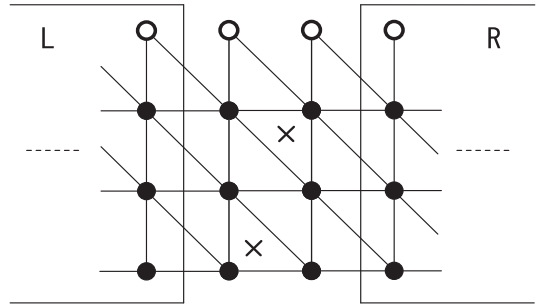} 
$$
\caption{The 3-leg ladder.
The correlation functions of the order parameters are obtained at 
the quantities on the positions denoted by crosses.}
\label{fig-Lattice-ladder3}
\end{figure}

\begin{figure}
$$
\epsfxsize=9cm \epsfysize=4.5cm \epsfbox{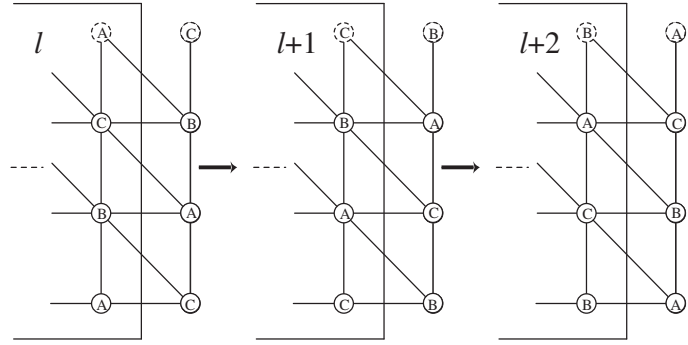} 
$$
\caption{Procedures of PWFRG}
\label{fig-PWFRG}
\end{figure}

In this section, we perform PWFRG on a $3\times\infty$ lattice with
the periodic boundary condition depicted in Fig.~\ref{fig-Lattice-ladder3}.
We expect that the model on this lattice reproduces well properties
of the model on the triangular lattice.
Since this lattice contains the
minimum length of the unit cell of the triangular lattice both in the
rung and leg directions.
The unit cell of the order parameter
is $3\times 3$, and therefore it had better take this unit cell as a unit of 
the renormalization process. However, it is difficult to
perform this process because of the limitation of the memory. Thus, we
take a $3\times 1$ cell as a unit cell of the RG process. 
In Fig.\ref{fig-PWFRG}, the procedure of PWFRG is illustrated. 
When we renormalize the block Hamiltonian by one step, the sublattice structure of the center spins shifts by one site along the leg direction.
In order to take account of this miss match of the periodicity, we set up three type of the block Hamiltonian. Namely, we perform almost the same computation three times and keep the consistency of the sublattice structure in the renormalization process by rotating the block Hamiltonians. 

\begin{figure}
$$\begin{array}{ccc}
\epsfxsize=5cm \epsfysize=4.0cm \epsfbox{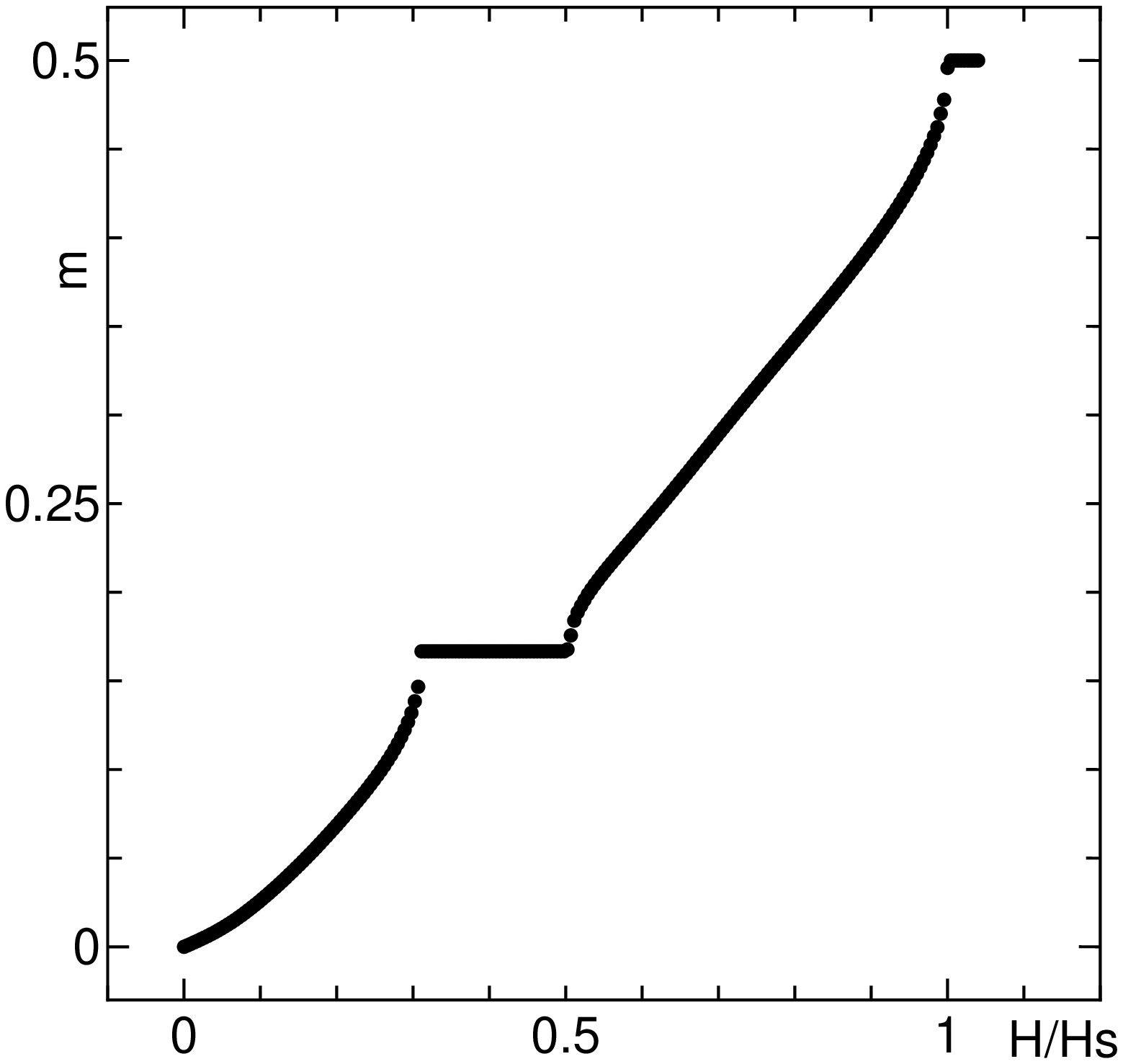}& 
\epsfxsize=5cm \epsfysize=4.0cm \epsfbox{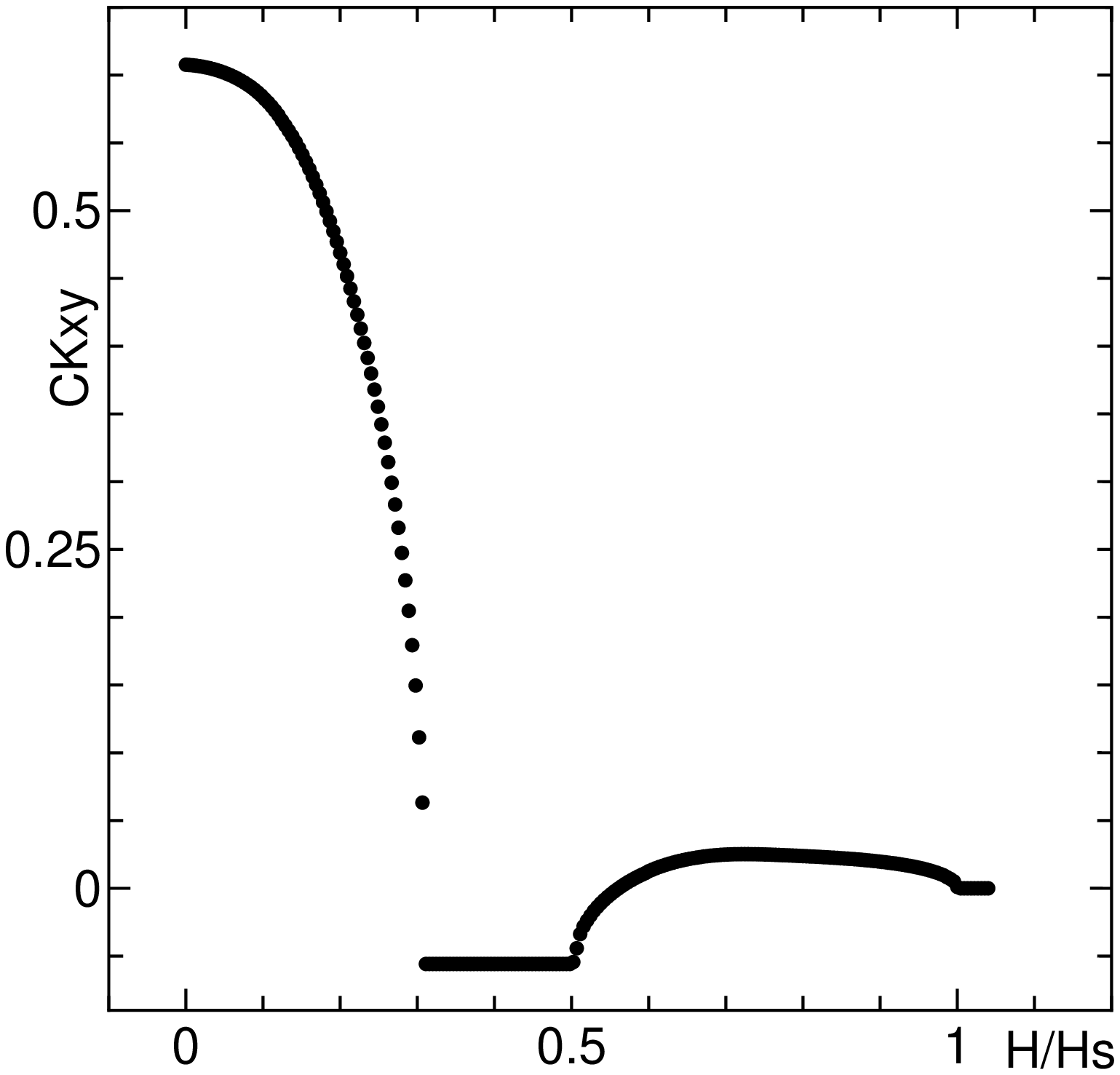} \\
{\rm (a)} & {\rm (b)} \\ 
\epsfxsize=5cm \epsfysize=4.0cm \epsfbox{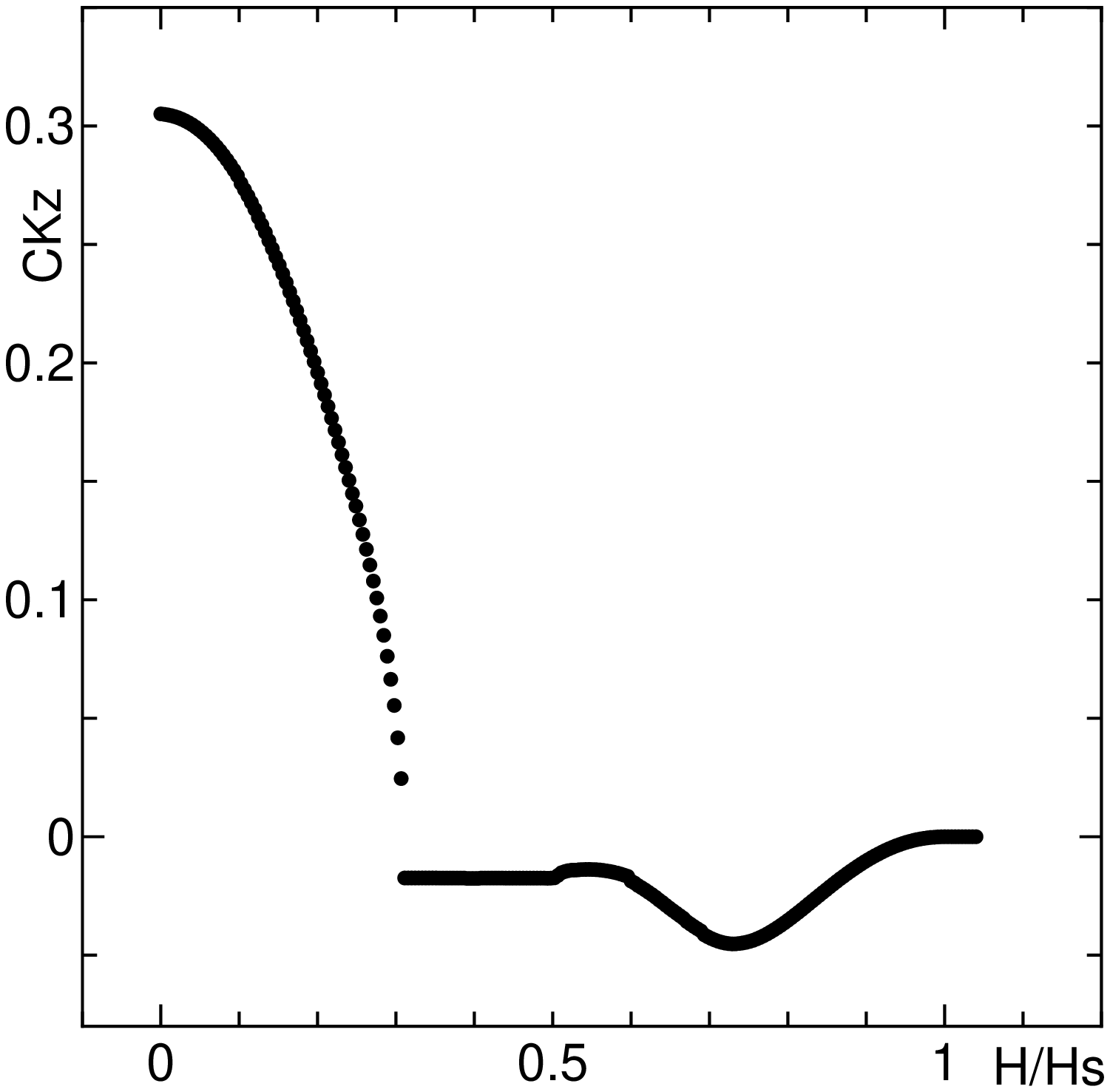}& 
\epsfxsize=5cm \epsfysize=4.0cm \epsfbox{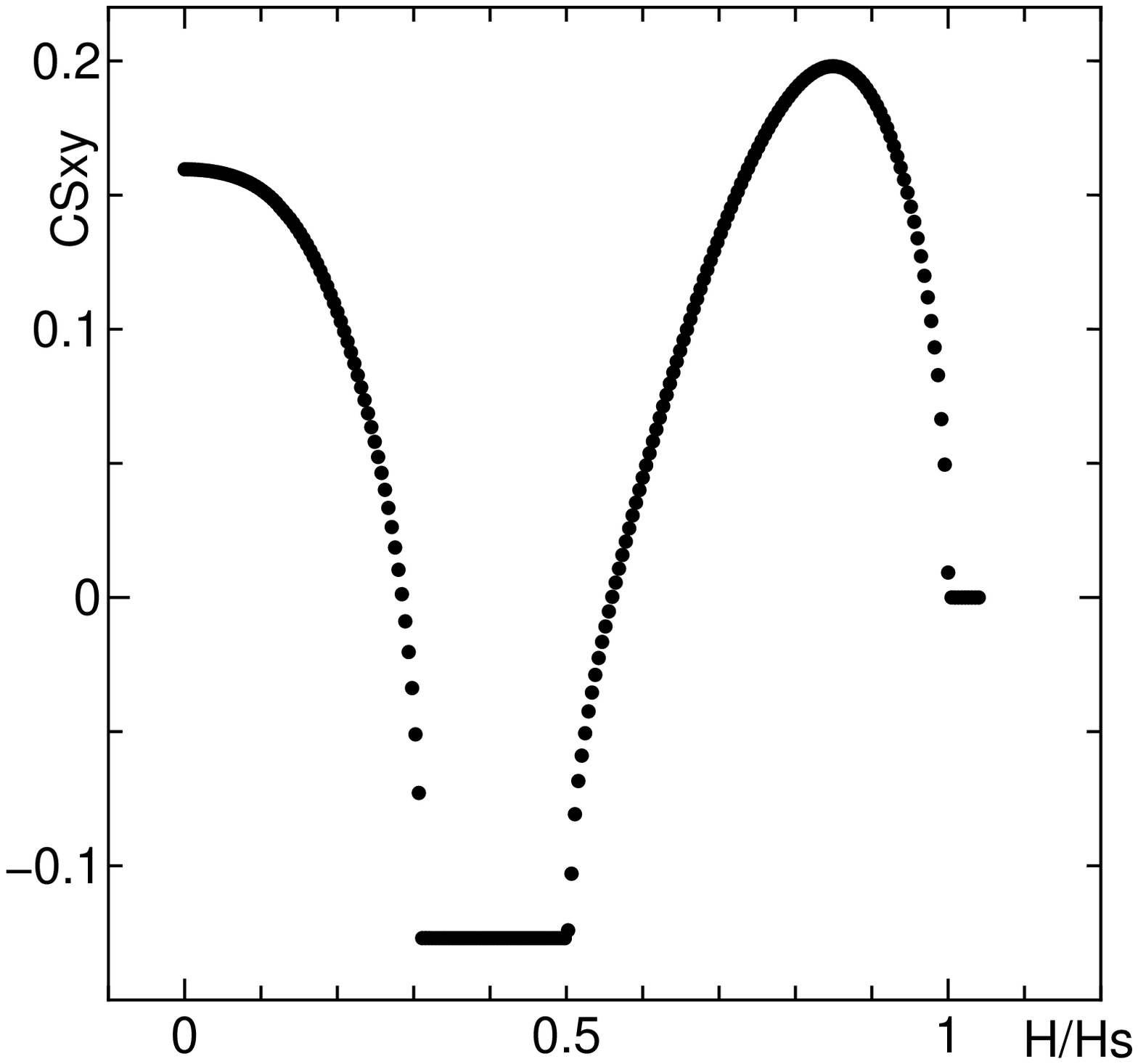} \\
{\rm (c)} & {\rm (d)} 
\end{array} $$
\caption{The field dependence of 
(a) the magnetization $m$, 
(b) $CK_{xy}$,
(c) $CK_z$, and
(d) $CS_{xy}$ 
as functions of the field  $H$ for the Heisenberg model $A=1.0$.}
\label{fig-data3-10}
\end{figure}
\begin{figure}
$$\begin{array}{ccc}
\epsfxsize=5cm \epsfysize=4.0cm \epsfbox{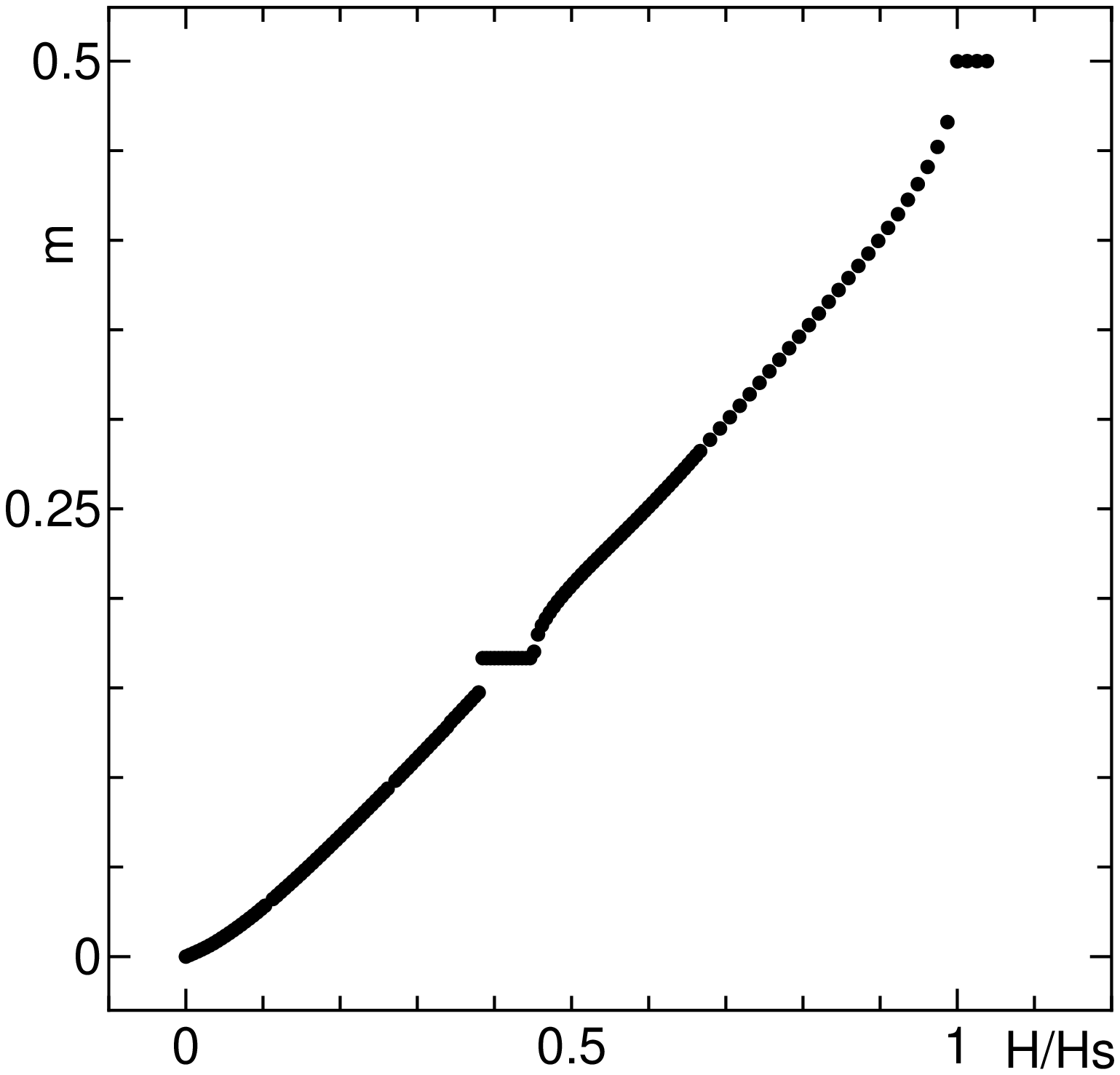}& 
\epsfxsize=5cm \epsfysize=4.0cm \epsfbox{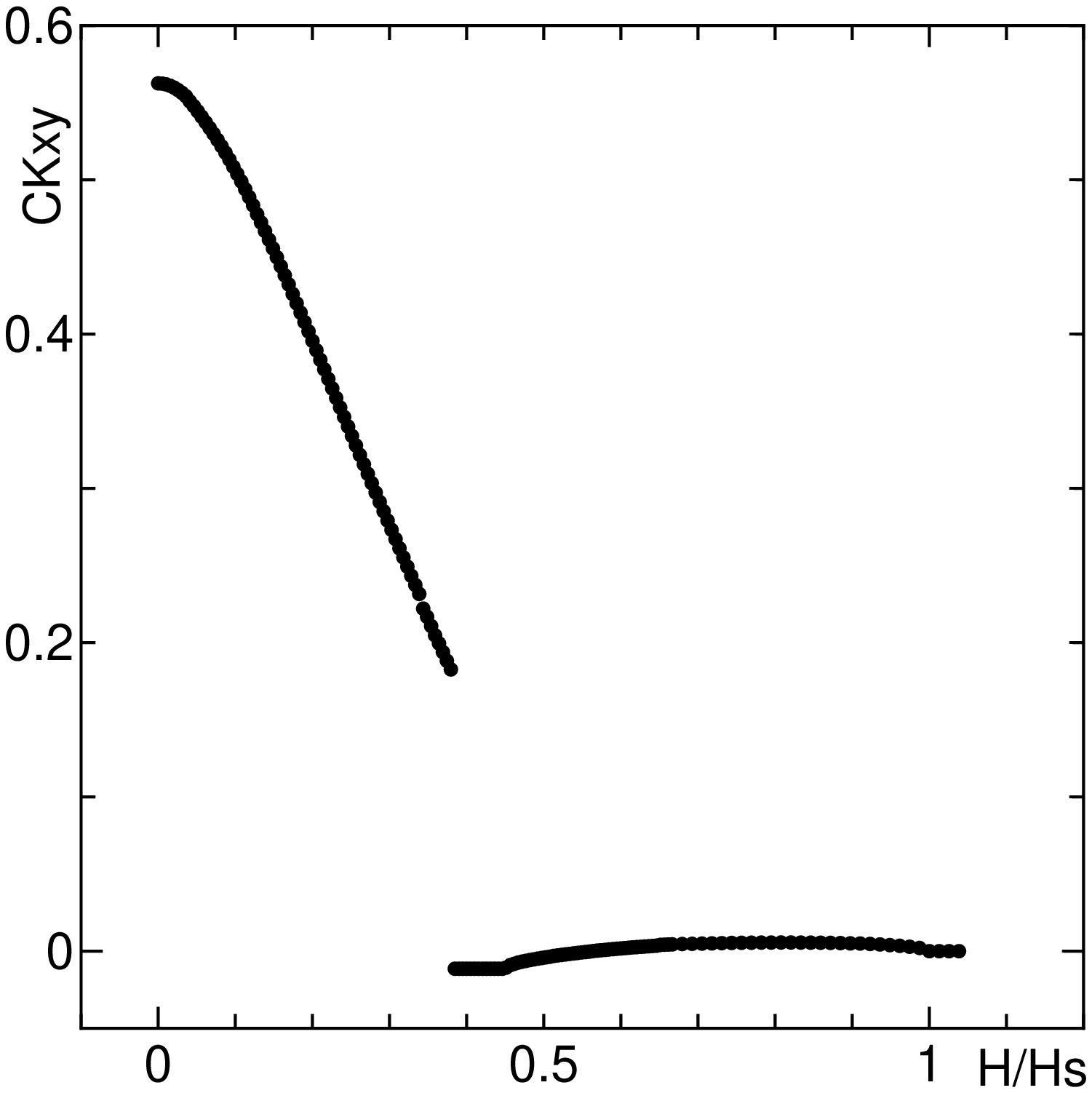} \\
{\rm (a)} & {\rm (b)} \\ 
\epsfxsize=5cm \epsfysize=4.0cm \epsfbox{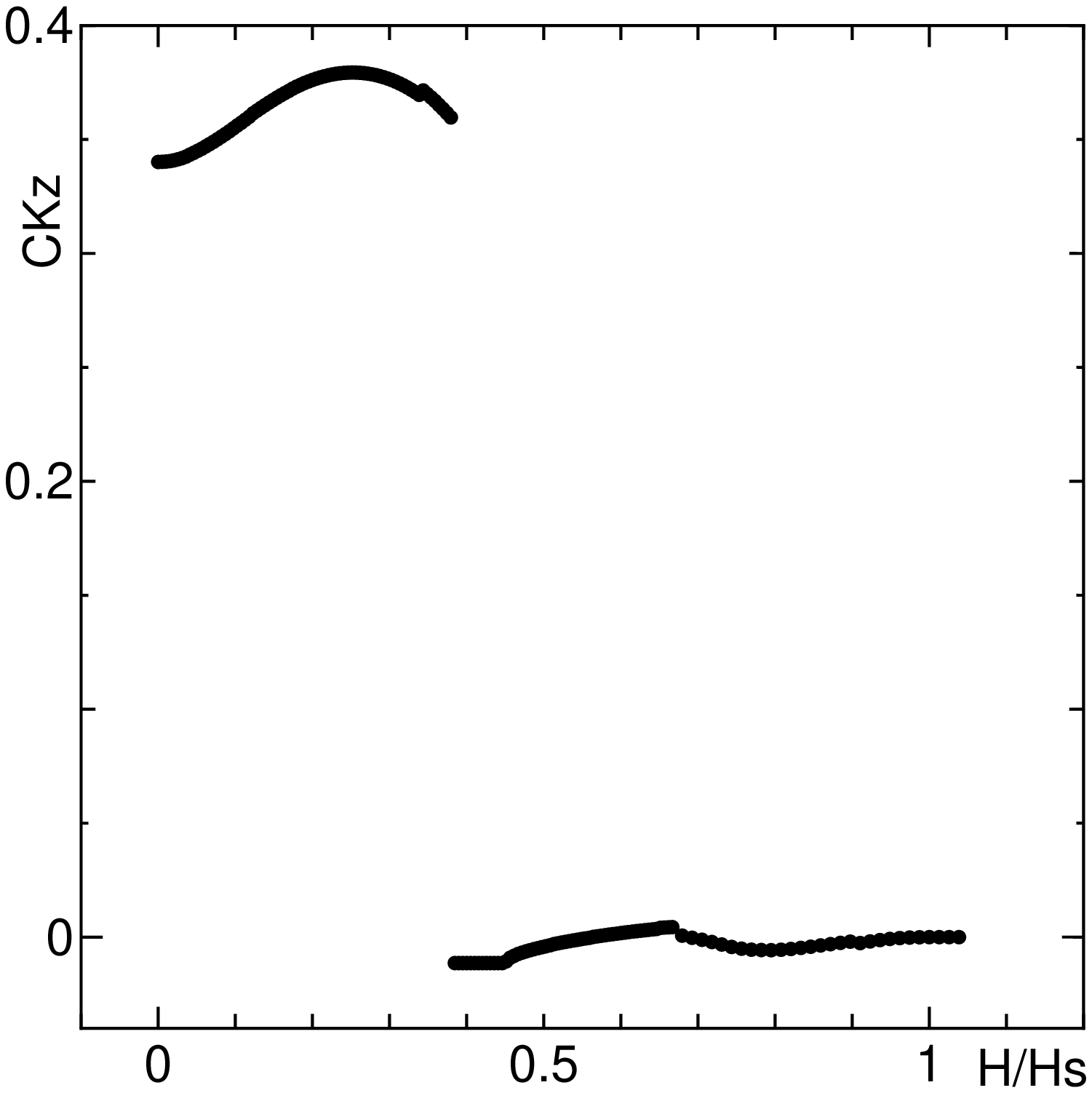}& 
\epsfxsize=5cm \epsfysize=4.0cm \epsfbox{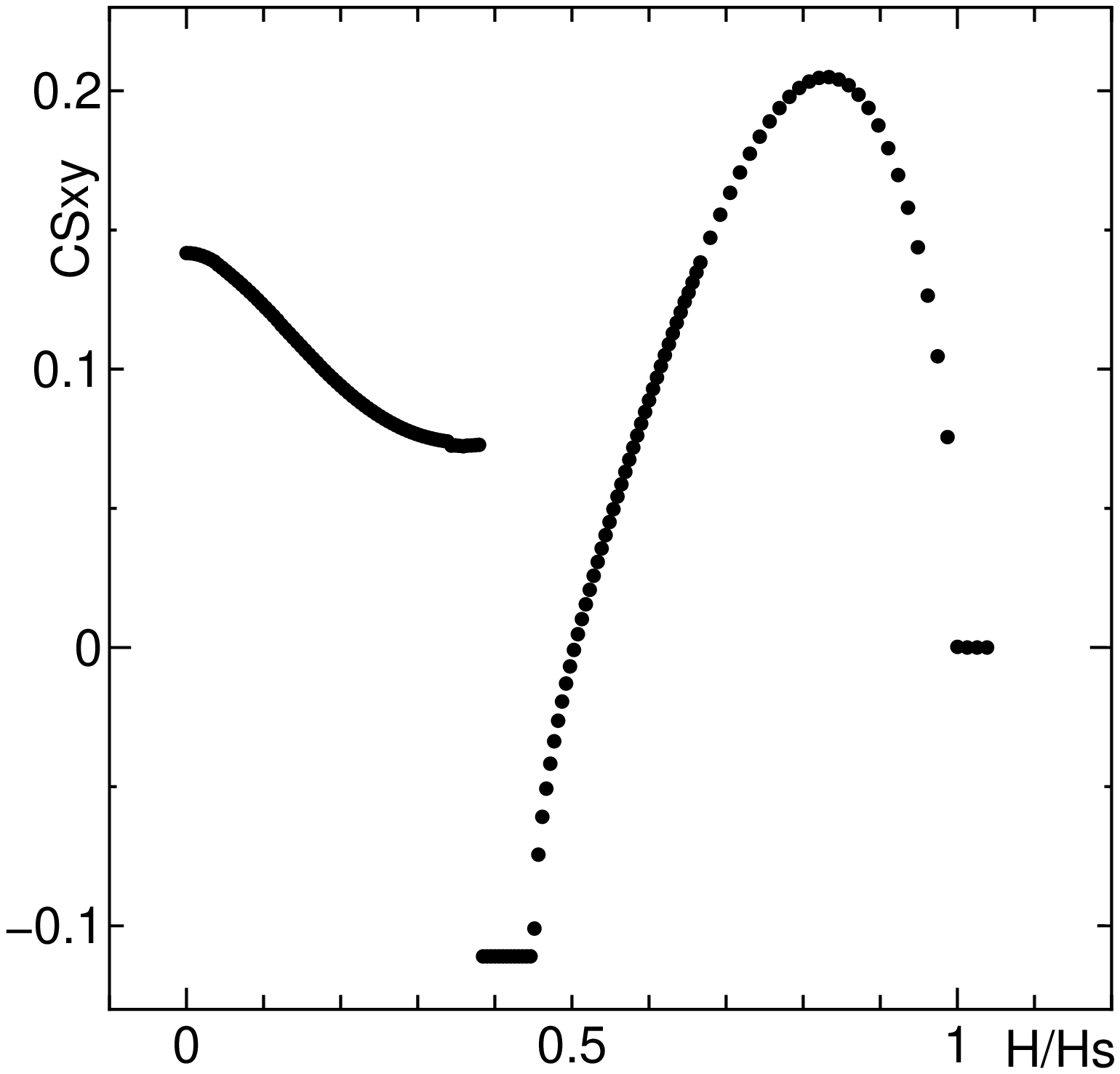} \\
{\rm (c)} & {\rm (d)} 
\end{array} $$
\caption{The field dependence of 
(a) the magnetization $m$, 
(b) $CK_{xy}$,
(c) $CK_z$, and
(d) $CS_{xy}$ 
as functions of the field  $H$ for the Heisenberg model $A=0.8$.}
\label{fig-data3-08}
\end{figure}
\begin{figure}
$$\begin{array}{ccc}
\epsfxsize=5cm \epsfysize=4.0cm \epsfbox{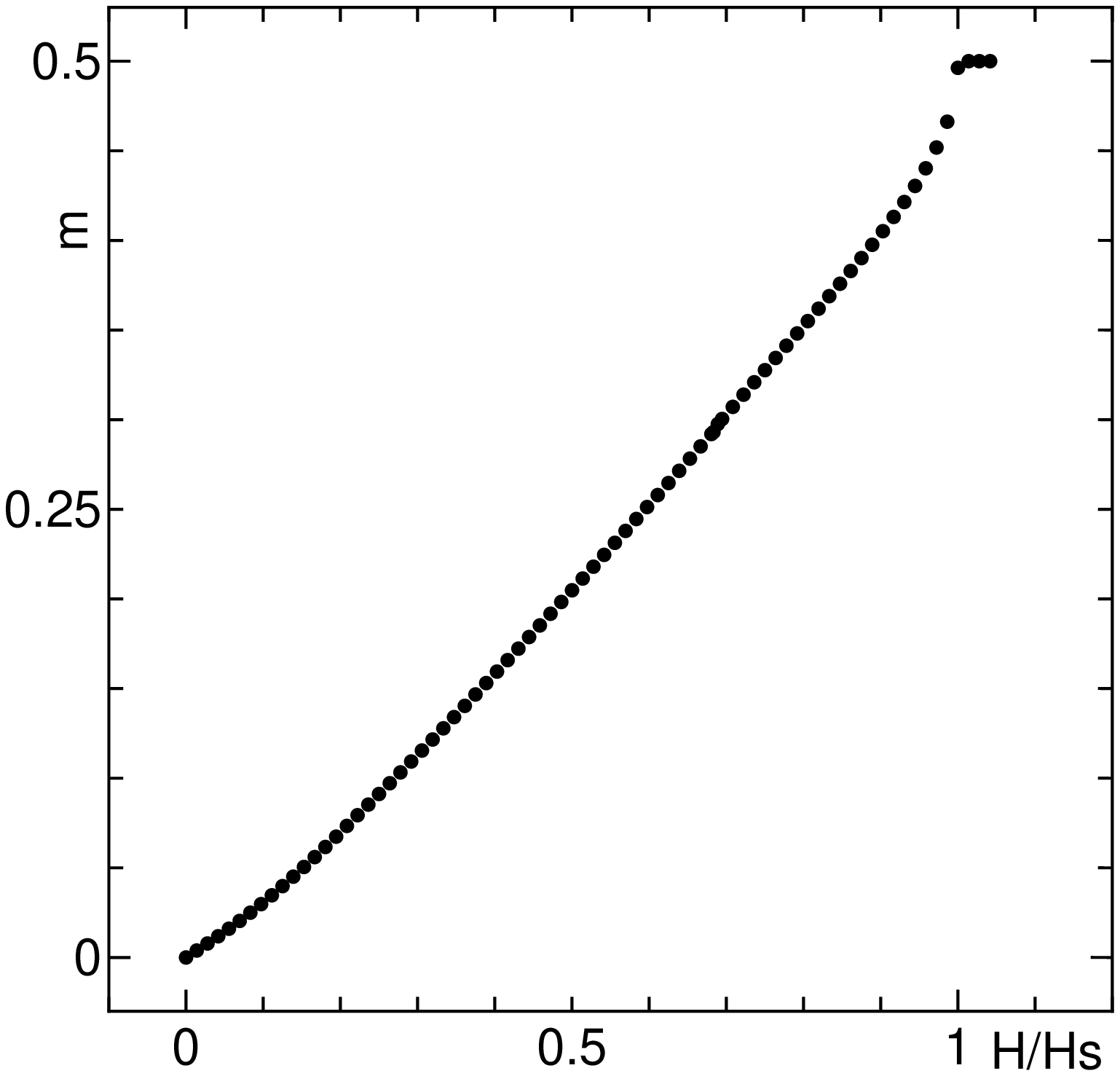}& 
\epsfxsize=5cm \epsfysize=4.0cm \epsfbox{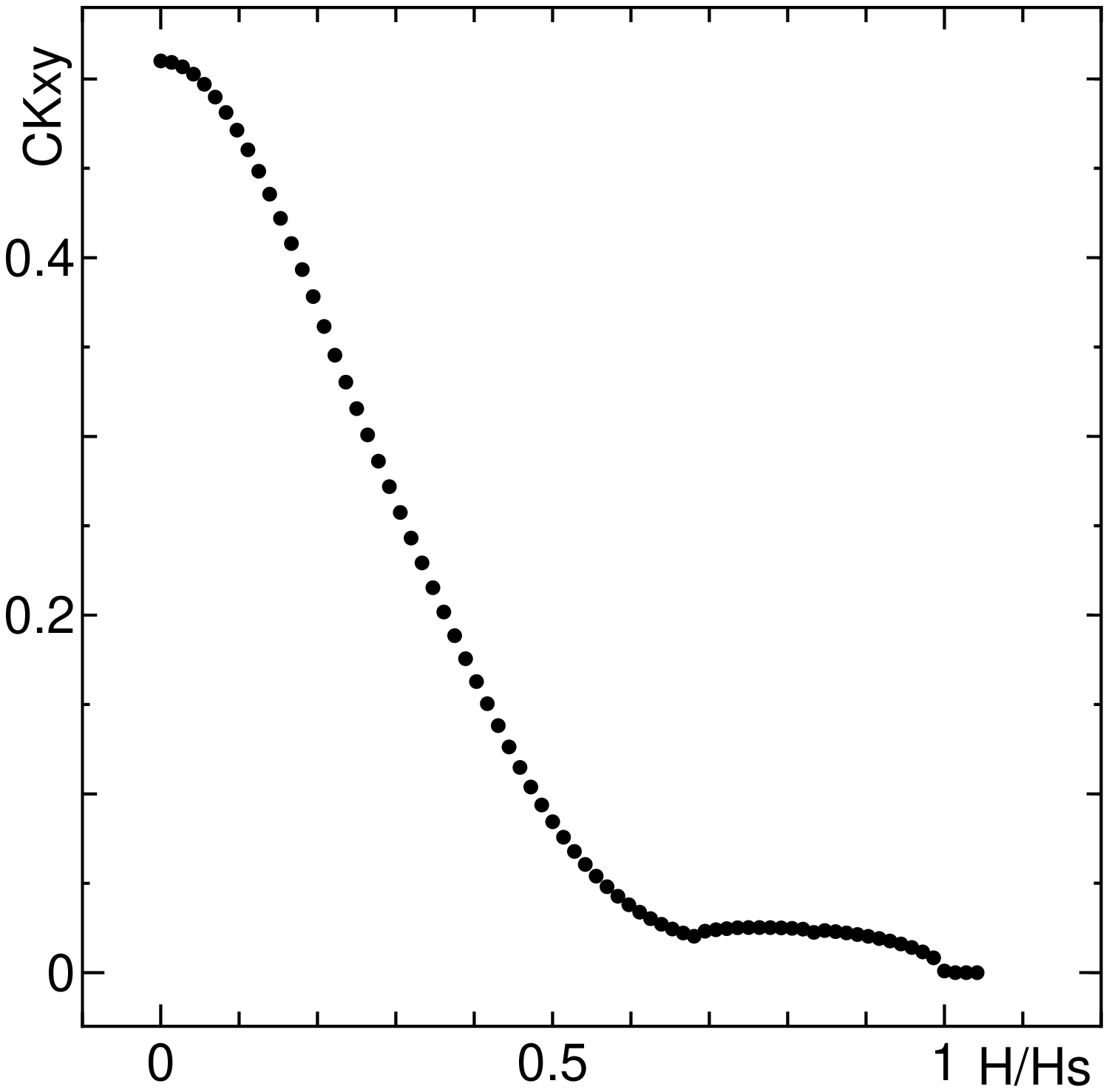} \\
{\rm (a)} & {\rm (b)} \\ 
\epsfxsize=5cm \epsfysize=4.0cm \epsfbox{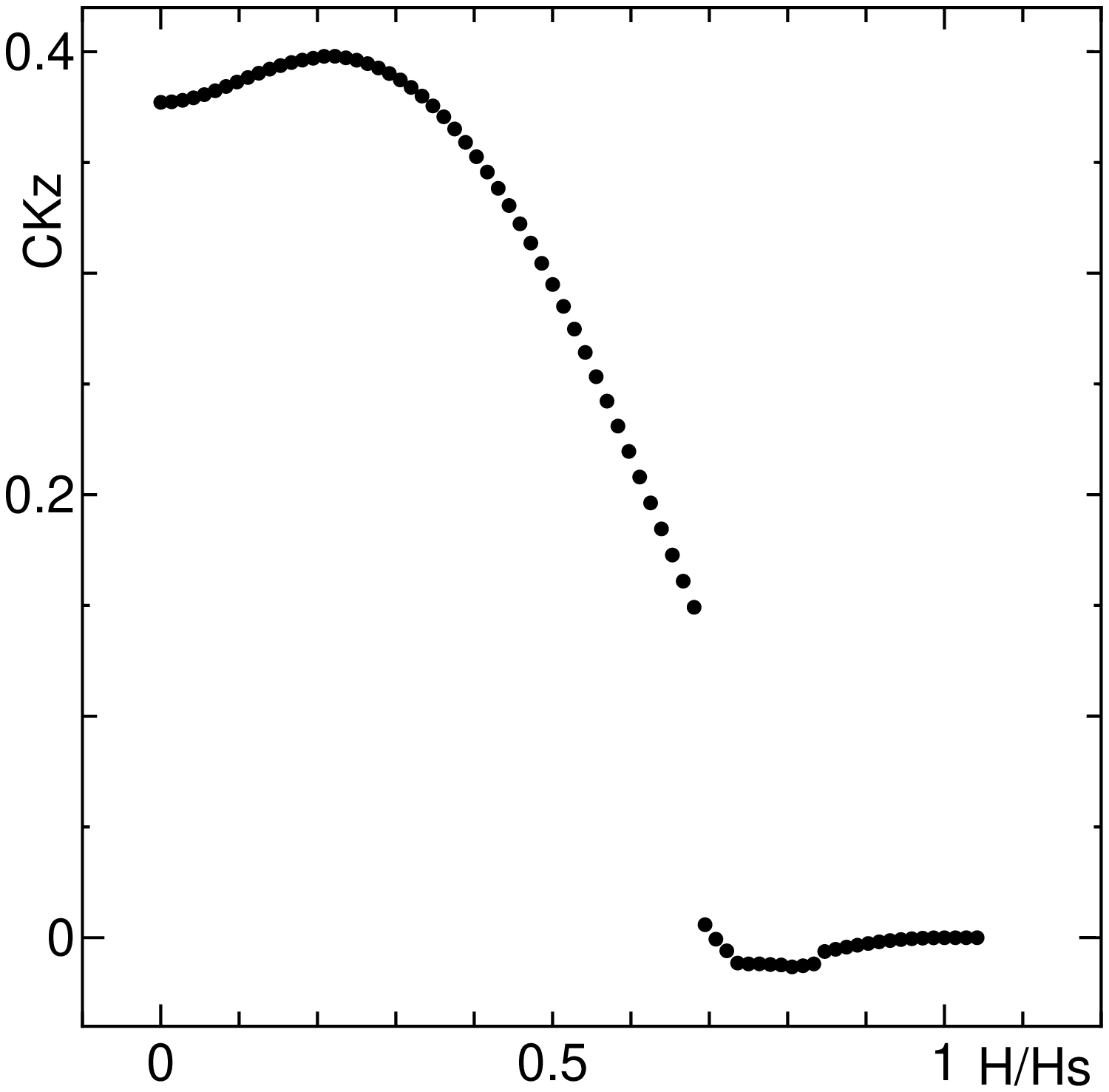}& 
\epsfxsize=5cm \epsfysize=4.0cm \epsfbox{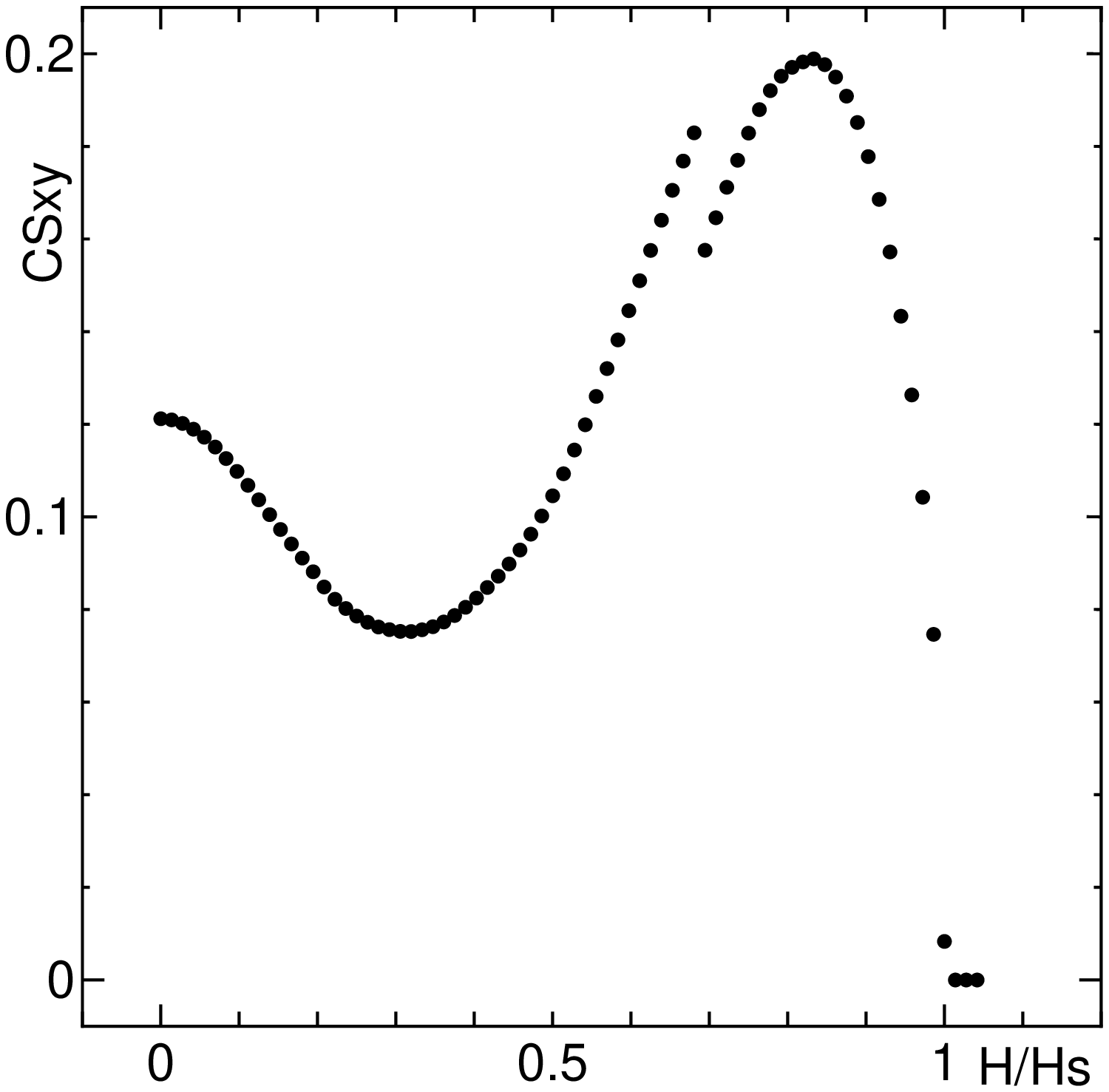} \\
{\rm (c)} & {\rm (d)} 
\end{array} $$
\caption{The field dependence of 
(a) the magnetization $m$, 
(b) $CK_{xy}$,
(c) $CK_z$, and
(d) $CS_{xy}$ 
as functions of the field  $H$ for the Heisenberg model $A=0.7$.}
\label{fig-data3-07}
\end{figure}
\begin{figure}
$$\begin{array}{ccc}
\epsfxsize=5cm \epsfysize=4.0cm \epsfbox{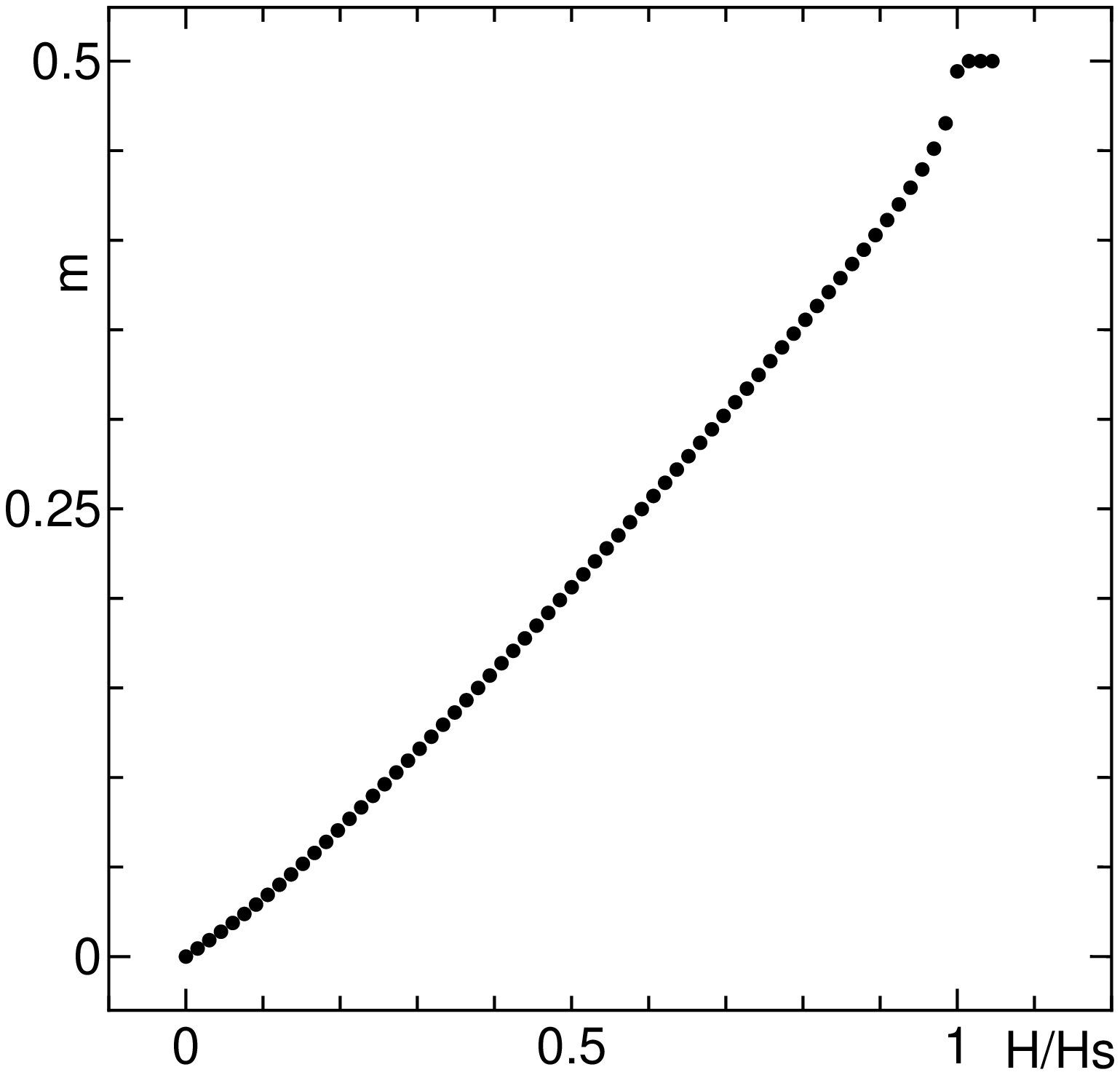}& 
\epsfxsize=5cm \epsfysize=4.0cm \epsfbox{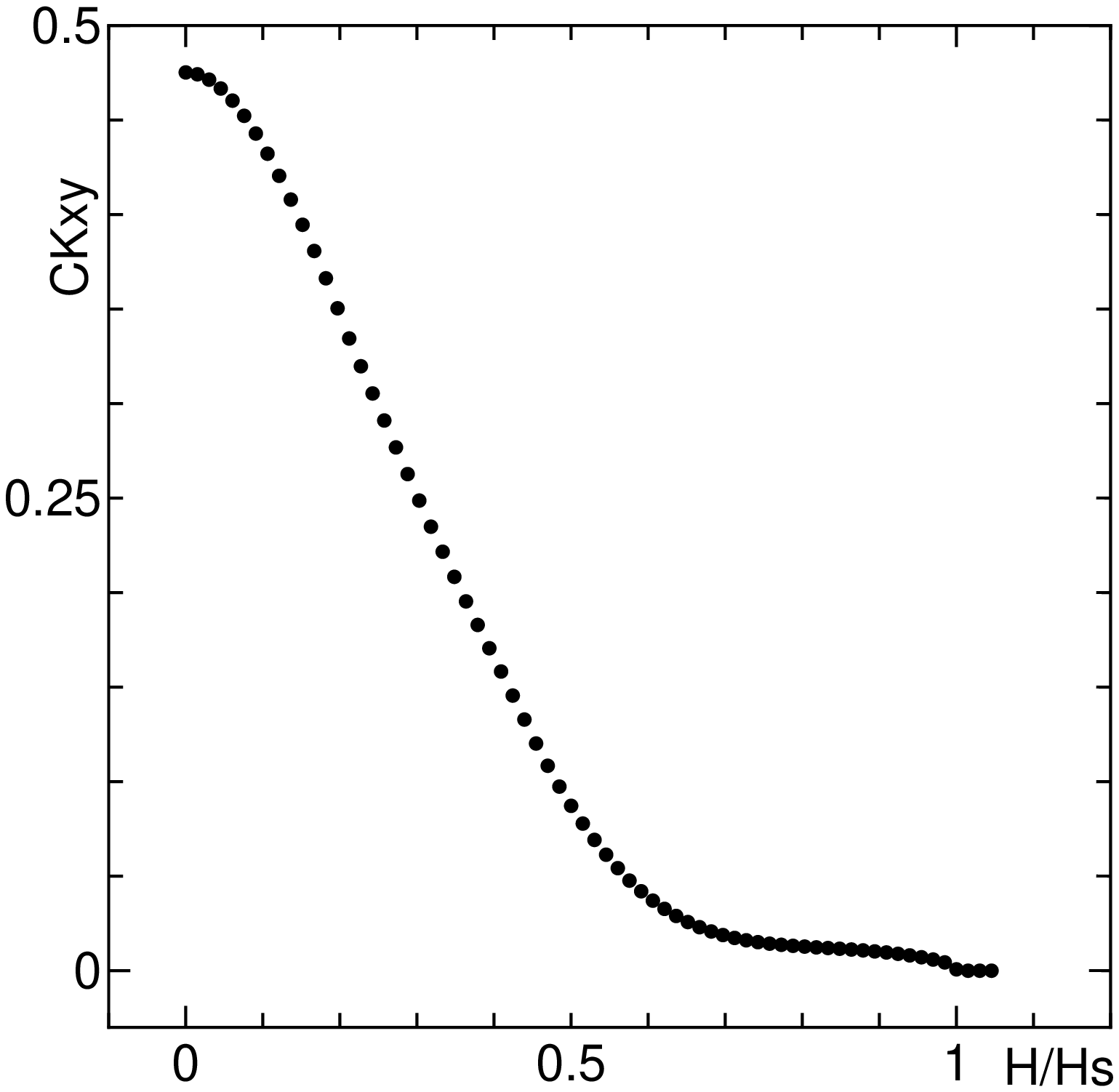} \\
{\rm (a)} & {\rm (b)} \\ 
\epsfxsize=5cm \epsfysize=4.0cm \epsfbox{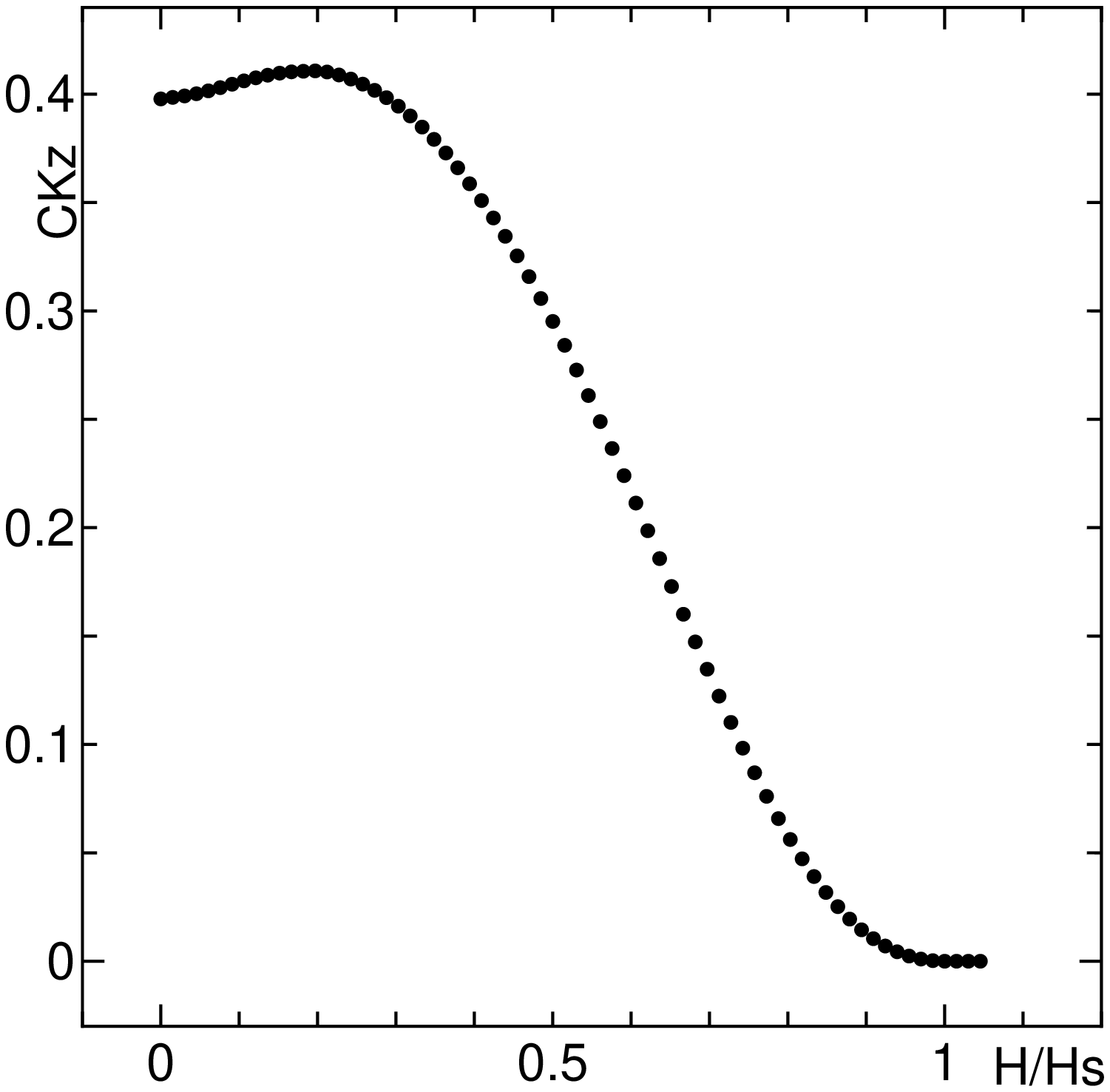}& 
\epsfxsize=5cm \epsfysize=4.0cm \epsfbox{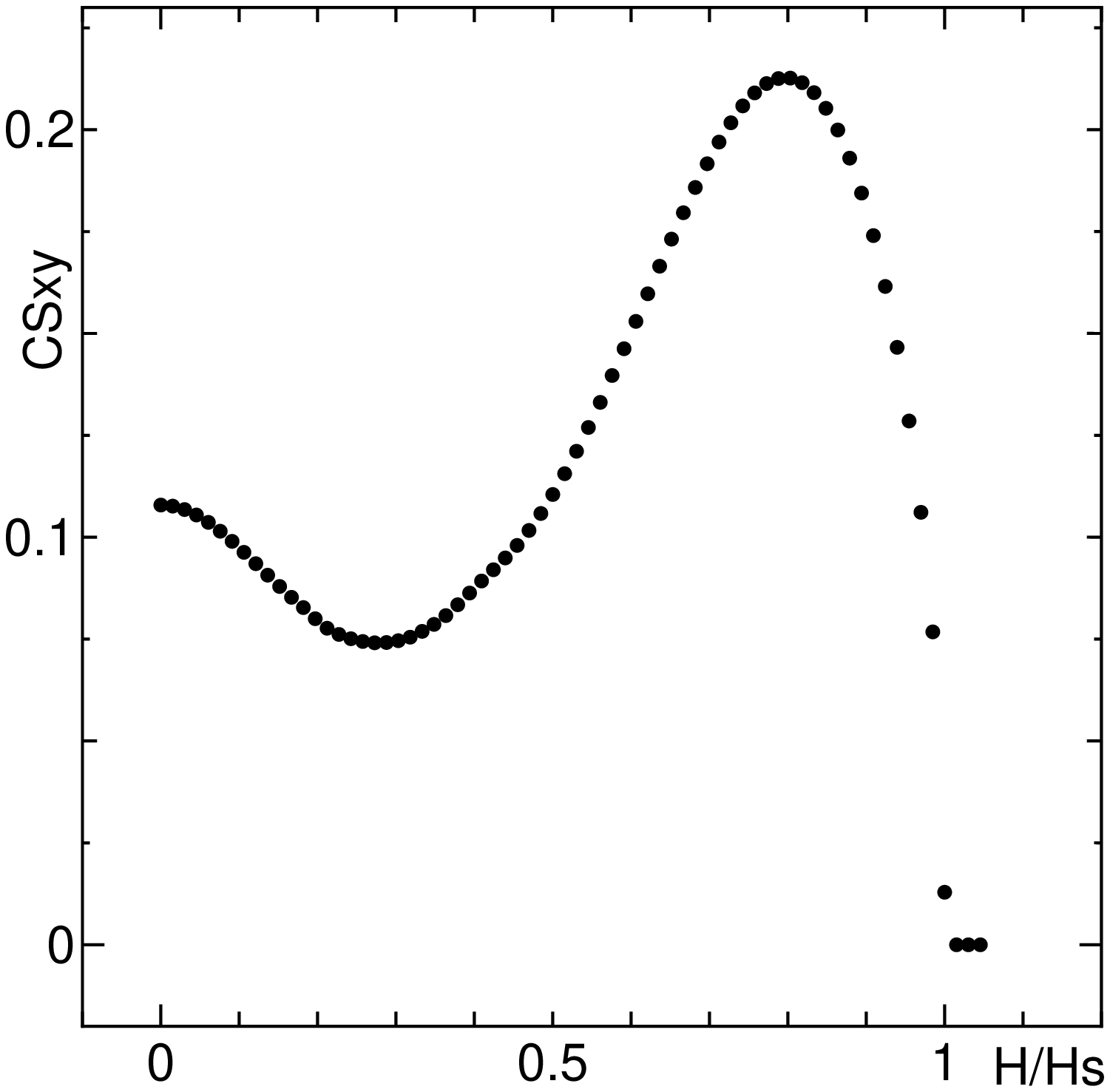} \\
{\rm (c)} & {\rm (d)} 
\end{array} $$
\caption{The field dependence of 
(a) the magnetization $m$, 
(b) $CK_{xy}$,
(c) $CK_z$, and
(d) $CS_{xy}$ 
as functions of the field  $H$ for the Heisenberg model $A=0.6$.}
\label{fig-data3-06}
\end{figure}

In Fig.\ref {fig-data3-10}, we depict the magnetization for $A=1.0$. Here, we find a magnetization with a plateau at $1/3$ of the full moment, which is similar to
what we found in the classical model.
Here we find that the $xy$-component of the chirality decreases with the field, and disappears at the 1/3 plateau.
When the field further increases, $CS_{xy}$ appears again. 
This observation is the same as those we found in the section \ref{ch:finite}.
Here we also find the relatively large $z$-component of the chirality. This may be due to the fact that we observe the local correlation function instead of the order parameter itself. 
When we reduce $A$ to 0.8 (see in Fig.\ref {fig-data3-08}), qualitatively similar behavior is observed. It should be noted that, however, the plateau width decreases, and the $z$-component of the
chirality increases.
For $A=0.7$, we find a low field phase with $z$-component of the
chirality and it changes discontinuously to a high field phase
of the non-collinear structure without chirality. (Fig.\ref {fig-data3-07})
This is similar to
what we found in the classical model with a weak XY anisotropy.
We also find that the plateau disappears and a jump appears instead of the plateau. (Fig.\ref{fig-data3-07}(a))
The magnitude of the anisotropy at which the plateau disappears obtained in this study 
is close to that obtained by A. Honecher et al.\cite{richter} ($A$ = 0.76 $\pm$ 0.03), 
which is the result of the exact diagonalization study up to 36 sites.
As the anisotropy becomes stronger, 
the umbrella structure is stabilized even in higher field region. 
For large anisotropy ($A=0.6$), we have no phase transition as the classical case.
(Fig.\ref {fig-data3-06})

\begin{figure}
$$\begin{array}{c}
\epsfxsize=5cm \epsfysize=5.0cm \epsfbox{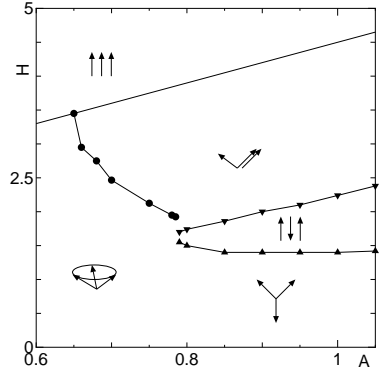} 
\end{array} $$
\caption{The ground state phase diagram in the $(A,H)$ coordinate.
The solid circles represent the boundary of the umbrella phase and 
non-collinear phase. The triangles represent the upper and lower edges 
of the 1/3 plateau.}
\label{fig-phase-diagram}
\end{figure}

Summarizing the dependence of the field-induced phases on the
degree of anisotropy,  
we illustrate in Fig.~\ref{fig-phase-diagram} a phase diagram in the $(A,H)$ coordinate.
In this figure, we find very similar structure to that found in the 
classical model.\cite{WMS}
The boundary between the umbrella phase to the Y-shape phase 
is not determined. For this determination we have to 
study which of $\kappa_z$ or $\kappa_{xy}$ dominates by
studying the distance dependence of the correlation functions,
although so far we obtained only correlation functions
between the triangles denoted by crosses in Fig.~\ref{fig-Lattice-ladder3}.

\section{Summary and Discussion}

 From the present observations we found that the ground state
of the $S=1/2$ XY-like Heisenberg antiferromagnetic model
depends on the anisotropy sensitively, where the ground state 
configuration is selected by the quantum correction.  
Then we found that the effect of quantum fluctuations
in the ground state is very similar to that of thermal fluctuations
at finite temperatures. 
In fact, the result for the three-leg ladder shows that  
the structure of the phase diagram (Fig.\ref{fig-phase-diagram}) 
is very similar to that of the corresponding classical model.
In the case of two-leg ladder, however, quantum fluctuations cause 
a special property which has no classical counter part.
This kind of change of the effect of quantum fluctuations
with the width of the lattice is also an interesting observation.
We note that the difference of the quantum and thermal fluctuations 
is an important future problem.

Recently, a material with spin-tube consisting of 
alternate triangles has been found ; it has the same lattice as the 3-leg ladder 
lattice studied in the present paper (Fig.\ref{fig-Lattice-ladder3}).\cite{Nojiri}
We expect that the properties found in the present paper,
such as the successive phase transitions and also their
dependence on the spin anisotropy, will be observed in the variety of such materials experimentally.   

This work is supported by the Grant-in-Aid from the Ministry of Education,
Culture, Sports, Science and Technology, and also by NAREGI Nanoscience Project, Ministry of Education, Culture, Sports, Science and Technology, Japan.
One of the authors(K.O) is supported by Grant for Promotion of Niigata University Research Projects.
The simulations have been carried out by using the computational facility of
the Super Computer Center of Institute for Solid State Physics, University of Tokyo.

\end{document}